\documentclass[aps,prd,showpacs,amsmath,amssymb,superscriptaddress,twocolumn]{revtex4}

\usepackage{graphicx}
\usepackage{dcolumn}
\usepackage{bm}
\usepackage{verbatim} 
\usepackage{mathrsfs}
\usepackage{comment}

\newcommand{\hw}{{\sc herwig}}
\newcommand{\py}{{\sc pythia}}

\def\met{\mathbin{E\mkern - 11mu/_T}}
\newcommand{\pt}{$p_{T}$}

\newcommand{\cteq}{{CTEQ5L}}
\newcommand{\MC}{MC}

\def\ie{\textit{i.e.,} }

\def\IE{\textit{i.e.,} }

\newcommand{\figurewidth}{0.73\hsize}  
\newcommand{\dfigurewidth}{0.95\hsize} 
\newcommand{\wfigurewidth}{1.00\hsize} 
\newcommand{\sfigurewidth}{0.95\hsize} 
\newcommand{\rfigurewidth}{1.00\hsize} 

\begin{document}

\preprint{CDF/ANAL/JET/CDFR/9297}
\preprint{PRD Draft Version 4.03}
 
\title{Measurement of the Inclusive Jet Cross Section
at the Fermilab Tevatron $p\bar p$ Collider
Using a Cone-Based Jet Algorithm}

\date{\today}

\affiliation{Institute of Physics, Academia Sinica, Taipei, Taiwan 11529, Republic of China} 
\affiliation{Argonne National Laboratory, Argonne, Illinois 60439} 
\affiliation{University of Athens, 157 71 Athens, Greece} 
\affiliation{Institut de Fisica d'Altes Energies, Universitat Autonoma de Barcelona, E-08193, Bellaterra (Barcelona), Spain} 
\affiliation{Baylor University, Waco, Texas  76798} 
\affiliation{Istituto Nazionale di Fisica Nucleare Bologna, $^t$University of Bologna, I-40127 Bologna, Italy} 
\affiliation{Brandeis University, Waltham, Massachusetts 02254} 
\affiliation{University of California, Davis, Davis, California  95616} 
\affiliation{University of California, Los Angeles, Los Angeles, California  90024} 
\affiliation{University of California, San Diego, La Jolla, California  92093} 
\affiliation{University of California, Santa Barbara, Santa Barbara, California 93106} 
\affiliation{Instituto de Fisica de Cantabria, CSIC-University of Cantabria, 39005 Santander, Spain} 
\affiliation{Carnegie Mellon University, Pittsburgh, PA  15213} 
\affiliation{Enrico Fermi Institute, University of Chicago, Chicago, Illinois 60637} 
\affiliation{Comenius University, 842 48 Bratislava, Slovakia; Institute of Experimental Physics, 040 01 Kosice, Slovakia} 
\affiliation{Joint Institute for Nuclear Research, RU-141980 Dubna, Russia} 
\affiliation{Duke University, Durham, North Carolina  27708} 
\affiliation{Fermi National Accelerator Laboratory, Batavia, Illinois 60510} 
\affiliation{University of Florida, Gainesville, Florida  32611} 
\affiliation{Laboratori Nazionali di Frascati, Istituto Nazionale di Fisica Nucleare, I-00044 Frascati, Italy} 
\affiliation{University of Geneva, CH-1211 Geneva 4, Switzerland} 
\affiliation{Glasgow University, Glasgow G12 8QQ, United Kingdom} 
\affiliation{Harvard University, Cambridge, Massachusetts 02138} 
\affiliation{Division of High Energy Physics, Department of Physics, University of Helsinki and Helsinki Institute of Physics, FIN-00014, Helsinki, Finland} 
\affiliation{University of Illinois, Urbana, Illinois 61801} 
\affiliation{The Johns Hopkins University, Baltimore, Maryland 21218} 
\affiliation{Institut f\"{u}r Experimentelle Kernphysik, Universit\"{a}t Karlsruhe, 76128 Karlsruhe, Germany} 
\affiliation{Center for High Energy Physics: Kyungpook National University, Daegu 702-701, Korea; Seoul National University, Seoul 151-742, Korea; Sungkyunkwan University, Suwon 440-746, Korea; Korea Institute of Science and Technology Information, Daejeon, 305-806, Korea; Chonnam National University, Gwangju, 500-757, Korea} 
\affiliation{Ernest Orlando Lawrence Berkeley National Laboratory, Berkeley, California 94720} 
\affiliation{University of Liverpool, Liverpool L69 7ZE, United Kingdom} 
\affiliation{University College London, London WC1E 6BT, United Kingdom} 
\affiliation{Centro de Investigaciones Energeticas Medioambientales y Tecnologicas, E-28040 Madrid, Spain} 
\affiliation{Massachusetts Institute of Technology, Cambridge, Massachusetts  02139} 
\affiliation{Institute of Particle Physics: McGill University, Montr\'{e}al, Canada H3A~2T8; and University of Toronto, Toronto, Canada M5S~1A7} 
\affiliation{University of Michigan, Ann Arbor, Michigan 48109} 
\affiliation{Michigan State University, East Lansing, Michigan  48824}
\affiliation{Institution for Theoretical and Experimental Physics, ITEP, Moscow 117259, Russia} 
\affiliation{University of New Mexico, Albuquerque, New Mexico 87131} 
\affiliation{Northwestern University, Evanston, Illinois  60208} 
\affiliation{The Ohio State University, Columbus, Ohio  43210} 
\affiliation{Okayama University, Okayama 700-8530, Japan} 
\affiliation{Osaka City University, Osaka 588, Japan} 
\affiliation{University of Oxford, Oxford OX1 3RH, United Kingdom} 
\affiliation{Istituto Nazionale di Fisica Nucleare, Sezione di Padova-Trento, $^u$University of Padova, I-35131 Padova, Italy} 
\affiliation{LPNHE, Universite Pierre et Marie Curie/IN2P3-CNRS, UMR7585, Paris, F-75252 France} 
\affiliation{University of Pennsylvania, Philadelphia, Pennsylvania 19104} 
\affiliation{Istituto Nazionale di Fisica Nucleare Pisa, $^q$University of Pisa, $^r$University of Siena and $^s$Scuola Normale Superiore, I-56127 Pisa, Italy} 
\affiliation{University of Pittsburgh, Pittsburgh, Pennsylvania 15260} 
\affiliation{Purdue University, West Lafayette, Indiana 47907} 
\affiliation{University of Rochester, Rochester, New York 14627} 
\affiliation{The Rockefeller University, New York, New York 10021} 

\affiliation{Istituto Nazionale di Fisica Nucleare, Sezione di Roma 1, $^v$Sapienza Universit\`{a} di Roma, I-00185 Roma, Italy} 

\affiliation{Rutgers University, Piscataway, New Jersey 08855} 
\affiliation{Texas A\&M University, College Station, Texas 77843} 
\affiliation{Istituto Nazionale di Fisica Nucleare Trieste/\ Udine, $^w$University of Trieste/\ Udine, Italy} 
\affiliation{University of Tsukuba, Tsukuba, Ibaraki 305, Japan} 
\affiliation{Tufts University, Medford, Massachusetts 02155} 
\affiliation{Waseda University, Tokyo 169, Japan} 
\affiliation{Wayne State University, Detroit, Michigan  48201} 
\affiliation{University of Wisconsin, Madison, Wisconsin 53706} 
\affiliation{Yale University, New Haven, Connecticut 06520} 
\author{T.~Aaltonen}
\affiliation{Division of High Energy Physics, Department of Physics, University of Helsinki and Helsinki Institute of Physics, FIN-00014, Helsinki, Finland}
\author{J.~Adelman}
\affiliation{Enrico Fermi Institute, University of Chicago, Chicago, Illinois 60637}
\author{T.~Akimoto}
\affiliation{University of Tsukuba, Tsukuba, Ibaraki 305, Japan}
\author{M.G.~Albrow}
\affiliation{Fermi National Accelerator Laboratory, Batavia, Illinois 60510}
\author{B.~\'{A}lvarez~Gonz\'{a}lez}
\affiliation{Instituto de Fisica de Cantabria, CSIC-University of Cantabria, 39005 Santander, Spain}
\author{S.~Amerio$^u$}
\affiliation{Istituto Nazionale di Fisica Nucleare, Sezione di Padova-Trento, $^u$University of Padova, I-35131 Padova, Italy} 

\author{D.~Amidei}
\affiliation{University of Michigan, Ann Arbor, Michigan 48109}
\author{A.~Anastassov}
\affiliation{Northwestern University, Evanston, Illinois  60208}
\author{A.~Annovi}
\affiliation{Laboratori Nazionali di Frascati, Istituto Nazionale di Fisica Nucleare, I-00044 Frascati, Italy}
\author{J.~Antos}
\affiliation{Comenius University, 842 48 Bratislava, Slovakia; Institute of Experimental Physics, 040 01 Kosice, Slovakia}
% ...

\collaboration{CDF Collaboration\footnote{With visitors from $^a$Universiteit Antwerpen, B-2610 Antwerp, Belgium, 
$^b$Chinese Academy of Sciences, Beijing 100864, China, 
$^c$University of Bristol, Bristol BS8 1TL, United Kingdom, 
$^d$University of California Irvine, Irvine, CA  92697, 
$^e$University of California Santa Cruz, Santa Cruz, CA  95064, 
$^f$Cornell University, Ithaca, NY  14853, 
$^g$University of Cyprus, Nicosia CY-1678, Cyprus, 
$^h$University College Dublin, Dublin 4, Ireland, 
$^i$University of Edinburgh, Edinburgh EH9 3JZ, United Kingdom, 
$^j$Universidad Iberoamericana, Mexico D.F., Mexico, 
$^k$University of Manchester, Manchester M13 9PL, England, 
$^l$Nagasaki Institute of Applied Science, Nagasaki, Japan, 
$^m$University de Oviedo, E-33007 Oviedo, Spain, 
$^n$Queen Mary, University of London, London, E1 4NS, England, 
$^o$Texas Tech University, Lubbock, TX  79409, 
$^p$IFIC(CSIC-Universitat de Valencia), 46071 Valencia, Spain,
$^x$Royal Society of Edinburgh/Scottish Executive Support Research Fellow, 
}}
\noaffiliation


\begin{abstract}

We present a measurement of the inclusive jet cross section 
in $p\bar p$ collisions at $\sqrt{s}=1.96$ TeV based on 
data collected by the CDF II detector with an
integrated luminosity of 1.13 fb$^{-1}$~\cite{erratum}.
The measurement was made using the cone-based Midpoint jet clustering
algorithm in the rapidity region of $|y|<2.1$. 
The results are consistent with next-to-leading-order perturbative QCD
predictions based on recent parton distribution functions (PDFs),
and are expected to provide increased precision in PDFs at high
parton momentum fraction $x$.
The results are also compared to the recent inclusive jet cross
section measurement using the
$k_T$ jet clustering algorithm, and we find that the
ratio of the cross sections measured with the two algorithms
is in agreement with theoretical expectations over a large range of
jet transverse momentum and rapidity.
\end{abstract}

\pacs{13.87.Ce, 12.38.Qk, 13.85.Ni}

\maketitle

\section{\label{sec:Intro}Introduction}

The measurement of the differential inclusive jet cross section
at the Fermilab Tevatron probes the highest momentum transfers in
particle collisions currently attainable in any accelerator
experiment, and thus is potentially sensitive to new physics such as
quark substructure~\cite{NewPhys1,NewPhys2}.
The measurement also provides a direct test of predictions of
perturbative quantum chromodynamics (pQCD)~\cite{pqcd1,pqcd2,qcdbook}.
The inclusive jet cross section measurements at Tevatron Run
II~\cite{Midpoint_RC,KT_PRL,KT_PRD,Dzero_Run2_incjet} cover 
up to 600 GeV/{\it c} in jet transverse momentum $p_T$~\cite{coordinate},
and range over more than eight orders of magnitude in differential
cross section.
Comparisons of the measured cross section with pQCD predictions
provide constraints on the parton distribution function (PDF) of the
(anti)proton, in particular at high momentum fraction $x$ ($x\gtrsim0.3$)
where the gluon distribution is poorly constrained~\cite{CTEQ6.1M}.
Further constraints on the gluon distribution at high $x$
will contribute to reduced uncertainties on theoretical
predictions of many interesting physics processes both for experiments at
the Tevatron and for future experiments at the Large Hadron Collider
(LHC).
One example is $t\bar{t}$ production at the
Tevatron for which the dominant PDF uncertainty arises from the
uncertainty in the high-$x$ gluon distribution.
In addition, searches for new physics beyond the standard model at high
$p_T$ such as quark substructure require precise knowledge
of PDFs at high $x$.

Jets are defined by algorithms which cluster together objects such as
energies measured in calorimeter towers, particles, or partons.
Jet clustering relies on the association of objects based 
either on proximity in coordinate space (as in cone algorithms) or in
momentum space (as in $k_T$
algorithms)~\cite{CDF_midpoint1,TeV4LHC,JetReview,LesHouches_tool}.
The CDF Collaboration recently published a measurement of the inclusive
jet cross section in the rapidity region
$0.1<|y|<0.7$~\cite{coordinate} using a cone-based jet clustering
algorithm~\cite{search_cone}
based on 0.39 $\mbox{fb}^{-1}$ of the Run II
data~\cite{Midpoint_RC}.
This paper presents an updated measurement based on 1.13
$\mbox{fb}^{-1}$ with the kinematic range
extended up to $|y|=2.1$, 
and comparisons with next-to-leading-order (NLO) pQCD predictions
based on recent PDFs of the proton~\cite{CTEQ6.1M,MRST2004}.
The extension of the rapidity range significantly increases the
kinematic reach in $x-Q$ space, where $Q$ denotes the momentum
transfer, and helps to further constrain the proton PDFs.
The D0 Collaboration also recently reported a measurement of
the inclusive jet cross section using 0.70 $\mbox{fb}^{-1}$ of data in
the rapidity region $|y|<2.4$~\cite{Dzero_Run2_incjet}.

Similar measurements of the inclusive jet cross section have been
made by the CDF Collaboration in Run II using the $k_T$ jet
clustering algorithm~\cite{KTalgo_Ellis} in the region of
$0.1<|y|<0.7$~\cite{KT_PRL} and later in the region up to
$|y|=2.1$~\cite{KT_PRD}.
The $k_T$ algorithm has been used successfully at $e^+e^-$ and
$e^{\pm}p$ collider experiments; however,
the cone algorithms have been used traditionally 
at hadron collider experiments, 
mainly due to the associated simplicity in constructing
corrections for the underlying event and for multiple interactions in
the same bunch crossing~\cite{CDF_midpoint1}.
It is worth noting that previous measurements made at the Tevatron
in Run I using a cone
algorithm and $k_T$ algorithm showed only marginal
agreement~\cite{incjet_run1_dzero,incjet_run1_kt_dzero}.

The rest of this paper proceeds as follows:
Section~\ref{sec:Detector} describes the CDF detector components most
relevant to this analysis. 
The details of the jet clustering algorithm and the data
sample used in this measurement are presented in
Secs.~\ref{sec:Clustering} and~\ref{sec:Data}.
Section~\ref{sec:Corrections} explains the methods used to correct the
CDF data for all detector effects, so that the measured cross section
may be directly compared to theoretical predictions.
The event samples from Monte Carlo (\MC) event generators
and CDF detector simulation that are used to derive these corrections
are also discussed in this section.
Systematic uncertainties in the cross section measurement are
discussed in Sec.~\ref{sec:Systematics}.
Section~\ref{sec:Theory} discusses NLO pQCD predictions on the
inclusive jet cross sections, and Sec.~\ref{sec:Results} presents
the measured cross sections and comparisons to those predictions. 
In Sec.~\ref{sec:CompareKt} the measured cross sections are also
compared to the recent measurement using the $k_T$ jet clustering
algorithm~\cite{KT_PRD}, and in Sec.~\ref{sec:Conclusions}
conclusions are presented.

\section{\label{sec:Detector}The CDF II Detector}

The CDF II detector, shown schematically in Fig.~\ref{fig:CDFIIdetector},
is described in detail elsewhere~\cite{CDF_detect3}.
Here, those components that are relevant to this measurement are
briefly described.
The central detector consists of a silicon vertex
detector (SVXII)~\cite{SVXII} and intermediate silicon layers
(ISL)~\cite{ISL}, covering the radial ranges of $1.5-11$ cm and
$19-30$ cm, respectively.
They are located inside a cylindrical open-cell drift chamber~\cite{COT} of 96 layers
organized in 8 superlayers with alternating structures of axial and 
$\pm2^{\circ}$ stereo readout within a radial range between 40 and 137
cm.
The tracking system is located inside a superconducting solenoid magnet which
provides an axial 1.4 T magnetic field.
Surrounding the magnet coil are projective-tower-geometry 
sampling calorimeters to measure the energy of interacting particles.

The central calorimeter covers the region of $|\eta|<1.1$ and is
divided into two halves at $|\eta|=0$.
It consists of 
48 modules, segmented into towers of granularity
$\Delta \eta \times \Delta\phi \approx  0.1 \times 0.26$.
The central electromagnetic calorimeter (CEM)~\cite{cem,cem_calib} consists of
lead-scintillator with a depth of about 18 radiation lengths;
the central hadron calorimeter (CHA)~\cite{cha_wha} consists of iron-scintillator
with a depth of approximately 4.7 interaction lengths.
The energy resolution of the CEM for electrons is
$\sigma(E_T) / E_T = 13.5\%/\sqrt{E_T ({\rm GeV})} \oplus 1.5\%$,
while the energy resolution of the CHA for charged pions that do not
interact in the CEM is
$\sigma(E_T) / E_T = 50\%/\sqrt{E_T ({\rm GeV})} \oplus 3\%$.

The forward region, $1.1 < |\eta| < 3.6$, is covered by the plug
calorimeters~\cite{pcal,pem_testbeam}
consisting of lead-scintillator for the electromagnetic section (PEM)
and iron-scintillator for the hadronic section (PHA).   
The PEM and PHA have a depth of about 23.2 radiation lengths
and 6.8 interaction lengths, respectively.
The PEM and PHA are identically segmented into 480 towers
of sizes which vary with $\eta$ ($\Delta \eta \times
 \Delta\phi \approx  0.1 \times 0.13$ at $|\eta|<1.8$ and 
 $\Delta \eta \times \Delta\phi \approx 0.6 \times 0.26$ at
 $|\eta|=3.6$). 
The energy resolution of the PEM for electrons is 
$\sigma(E_T)/E_T = 14\%/\sqrt{E_T ({\rm GeV})} \oplus 1\%$,
while the energy resolution of the PHA for charged pions that do not
interact in the PEM is
$\sigma(E_T)/E_T = 74\%\sqrt{E_T ({\rm GeV})} \oplus 4\%$.
The gap in the projective tower geometry between CHA and PHA, 
corresponding to $0.7 < |\eta| <1.3$,
is covered by an iron-scintillator endwall hadron calorimeter
(WHA)~\cite{cha_wha}
with segmentation similar to that of the central calorimeter. 
The WHA has a depth of approximately 4.5 interaction lengths, and a
resolution of 
$\sigma(E_T)/E_T = 75\%/\sqrt{E_T({\rm GeV})} \oplus 4\%$
for charged pions that do not interact in the electromagnetic section.

A system of Cherenkov luminosity counters (CLC)~\cite{lumi} is located
around the beampipe and inside the plug calorimeters.
The CLC detector, covering the range $3.6<|\eta|<4.6$, consists of two
modules on the two sides of the interaction region.
Each module consists of 48 thin and long gas Cherenkov counters
arranged in three concentric layers of 16 counters.
The CLC detector is used to measure the number of inelastic $p\bar{p}$
collisions per bunch crossing and thereby the luminosity.

\section{\label{sec:Clustering} Jet Clustering}

The definition of a jet is a fundamental step in the measurement of the
inclusive jet cross section.
Jets are collimated sprays of particles originating from quark or gluon
fragmentation. 
They must be defined by clustering algorithms, and the algorithms
are designed such that
the jets clustered from the complex structure of objects
(such as energies measured in calorimeter towers) in each event
represent the physical properties of the partons from the
hard scattering.
The commonly-used jet clustering algorithms can be categorized
into two classes, {\it i.e.}, cone-based algorithms and $k_T$
algorithms.
The two categories of algorithms have different strengths and
weaknesses in regards to comparisons between data and theoretical
predictions.
For example, as mentioned previously, the underlying event and
multiple interaction corrections are simpler for cone algorithms,
while $k_T$ algorithms have a smaller sensitivity to higher order
perturbative QCD
effects~\cite{CDF_midpoint1,JetReview,LesHouches_tool}.

The $k_T$ algorithms are based on pair-wise successive combinations.
In the $k_T$ algorithm~\cite{KTalgo_Ellis},
initially each object to be clustered is considered as a proto-jet,
and the quantities
$k_{T,i}^{2}       = p_{T,i}^{2}$ and
$k_{T,(i,j)}^{2}   =\min(p_{T,i}^{2},p_{T,j}^{2})\Delta R_{i,j}^{2}/D^{2}$
are then computed for each proto-jet and each pair of proto-jets,
respectively, where $p_{T,i}$ is the $p_T$ of the $i$-th
proto-jet, $\Delta R_{i.j}$ is the distance in a specified coordinate
space ({\it e.g.}, $y-\phi$ space)
between each pair of proto-jets, and $D$ is the parameter that
controls the size of the jet. 
If the smallest of these quantities is a $k_{T,i}^{2}$, 
that proto-jet becomes a jet and is removed from the list of
proto-jets, and if the smallest quantity is a $k_{T,(i,j)}^{2}$, 
the two proto-jets are merged into a single proto-jet and the original
two proto-jets are removed from the proto-jet list.
This process is iterated until all the proto-jets become jets.

In cone algorithms, objects in a cone in a specified coordinate space
are clustered, and the axis of the cone is
required to coincide with the direction of the cone defined by a sum
of all objects inside the cone. Such cones are referred to as stable
cones, and jets are formed from these stable cones.
Cone algorithms used in experiments so far search for stable cones
only from the locations of seeds, objects above a threshold,
in order to keep the CPU running time manageable.
The use of seeds makes these cone algorithms sensitive to soft
particles, and it has been pointed out that 
pQCD calculations with
cone algorithms used previously in Run I~\cite{incjet_run1_dzero,incjet_run1}
may face difficulties due to the presence of infrared
singularities~\cite{CDF_midpoint1,Giele_Kilgore}.
The Midpoint cone-based jet clustering algorithm~\cite{CDF_midpoint1}
used in this paper and also in other recent
measurements~\cite{Midpoint_RC,Dzero_Run2_incjet}
reduces this problem by placing additional seeds at the midpoint
between stable cones having a separation of less than twice the
clustering cone radius.
More details of the Midpoint jet clustering algorithm used in this
measurement are described below.

First, a list of objects to be clustered needs to be identified.
In this paper, jets are clustered at three different levels.
The list of objects to be clustered is different in each case:
\begin{itemize}
\item[(a)] Detector level (in data or MC events after detector simulation):
  CDF reconstructs jets from four-vectors associated with calorimeter towers.
  The four-vector associated with each tower is defined as a sum of vectors
  for the electromagnetic and hadronic sections.
  The vector of each section is defined as a massless
  four-vector with magnitude equal to the deposited energy and with
  direction from the primary interaction event vertex to the center of
  each section.
  To reduce the effect of detector noise, only towers with
  $p_T>100~\mbox{MeV}/c$ are included in the list.
\item[(b)] Hadron level (in MC events): four-vectors of the stable
   particles (mostly hadrons and photons from $\pi^0$
   decays)~\cite{hadron_level}
   are the basic elements to be
   clustered.
 \item[(c)] Parton level (in MC events or NLO pQCD theory):
   four-vectors of partons are used to form parton-level jets.
   In MC events, the partons before hadronization are used, and
   in the pQCD theory, the partons after all QCD radiation
   are used.
   There are at most three partons in the list in
   the NLO pQCD theory used in this paper.
\end{itemize}

Then, a list of seed objects is made
with the 
requirement that the \pt\ of the object 
exceeds a fixed threshold, which is set to $1~\mbox{GeV}/c$ in this
analysis.
At each seed location a cone of radius $R_{\rm cone}=0.7$ in $y-\phi$
space is constructed, and
the four-momentum vectors of all objects located in the cone are
summed.
This four-vector sum defines the \textit{centroid} of the cluster,
{\it i.e.},
\begin{eqnarray}
p^{\rm cluster}    &=&(E^{\rm cluster},{\bf p}^{\rm
  cluster})=\sum_{i\in {\rm cone}}(E^i,p_x^i, p_y^i, p_z^i ),  \nonumber \\
p_{T}^{\rm cluster}&=&\sqrt{{(p_x^{\rm cluster}})^2 + ({p_y^{\rm cluster}})^2},                    \nonumber \\
y^{\rm cluster}    &=&\frac{1}{2}\ln \left (
                  \frac{E^{\rm cluster}+p_z^{\rm cluster}}{E^{\rm cluster}-p_z^{\rm cluster}}\right ), \nonumber \\
\phi^{\rm cluster} &=&\tan^{-1}\left(p_y^{\rm cluster}/p_x^{\rm cluster}\right).
\label{eqn:4momentum-recombination}
\end{eqnarray}
This definition for the kinematics of a cluster is referred to as the 
four-vector recombination scheme~\cite{CDF_midpoint1}.
The four-vector of the cluster is then used as a new cone axis.
From this axis a new cone is drawn and the process of summing up the
four-vectors of all objects in the cone is repeated. 
This process is iterated until the cone axis and the centroid
coincide, indicating that a stable cone has been formed.

In the next step in the algorithm,
additional seeds are added at the midpoints between all pairs of
stable cones whose separation in $y-\phi$ space is less than $2R_{\rm
  cone}$.
A cone of radius $R_{\rm cone}$ is then drawn around the midpoint seed
and is used to form a stable cone.
If the resulting cone is not already in the list of stable cones,
it is added to the list. After all midpoint seeds have been explored,
the list of stable cones is complete.
As mentioned earlier, 
the use of these additional seeds reduces the sensitivity of the
algorithm to soft particles and makes this algorithm infrared safe up to
NLO in pQCD for inclusive jet cross sections.

It is possible that stable cones overlap, \IE an object
may be contained in more than one stable cone. To resolve these
configurations, a split-and-merge algorithm is employed.
After stable cones are sorted in decreasing \pt, 
overlapping stable cones are merged
if the \pt\ of the four-vector sum of shared objects between two
overlapping cones is more than a fraction, $f_{\rm merge}=0.75$, of
the $p_T$ of the lower-$p_T$ cone; otherwise,
the shared objects are assigned to the cone closer in $y-\phi$ space.
After cone overlaps are resolved and all objects are uniquely assigned
to a cluster, the resulting clusters are promoted to jets and 
their kinematic properties are determined using the four-vector
recombination scheme as defined in
Eq.~(\ref{eqn:4momentum-recombination}) where
the sum is over all objects assigned to the jet.
The Midpoint algorithm used in this measurement may then be summarized
as follows:
\begin{itemize}
\item[(1)] A list of seeds which includes only objects with $p_T>1$
  GeV/{\it c} is made.
\item[(2)] Stable cones with radius $R_{\rm cone}=0.7$ are constructed around
  each seed.
\item[(3)] An additional seed is added at the midpoint between each pair of
  stable cones separated by less than $2R_{\rm cone}$.  Each
  additional seed is used to search for stable cones that have
  not already been found.
\item[(4)] The stable cones are \pt-ordered and the split-and-merge
  procedure is performed to resolve overlapping cones.
\end{itemize}

\section{\label{sec:Data}Data Sample and Event Selection}

The measurement described in this paper is based on the data taken
from February 2002 until
February 2006 corresponding to an integrated luminosity of
$1.13\pm0.07$ $\mbox{fb}^{-1}$.
The data samples used in this measurement were collected using four
paths in the CDF three-level trigger system.
The level-1 trigger requires a
calorimeter trigger tower, consisting of a specific pair of
calorimeter towers adjacent in $\eta$,
to have $E_T>5$, $5$, $10$, and $10$ GeV in the four trigger paths,
respectively, for most of the time; however, the
$E_T$ threshold for the last path is changed from $10$ to $20$ GeV
in the course of the data-taking period in order to accomodate the
increase in the trigger rate due to increasing instantaneous
luminosity.
At level 2, the calorimeter towers are clustered using a
nearest-neighbor algorithm.
Events are required to have at least one level-2 trigger cluster with
$E_T >$ $15$, $40$, $60$, and $90$ GeV
in each of four trigger paths, respectively.
Events in these four paths are further required to have at least one
jet with $E_T >20$, $50$, $70$, and $100$ GeV at level 3, where the
jet clustering is performed using the CDF Run I cone algorithm with a
cone radius $R_{\rm cone}=0.7$~\cite{incjet_run1}.
These four jet trigger paths are referred to as ``jet20'', 
``jet50'', ``jet70'' and ``jet100'' hereafter.
The minimum $p_T$ at which jets from a given trigger path are used 
is determined by requiring a trigger efficiency greater than 99.5\%.
The trigger efficiencies in the region $0.1<|y|<0.7$ are
shown in Fig.~\ref{fig:trig_eff}. 

\begin{table*}[tp]
\caption{Summary of the jet triggers used in this measurement. For each
  dataset, the $E_T$ threshold on the trigger towers at level 1,
  calorimeter clusters at level 2 and jets clustered at level 3,
  and the corresponding prescale factors are shown. When multiple numbers are
  shown in a single column, it means the prescale factor or the $E_T$
  threshold changed during the data taking period studied.}
\begin{ruledtabular}
\begin{tabular}{lccccccc} 
Dataset &\multicolumn{2}{c}{level-1}&\multicolumn{2}{c}{level-2}&\multicolumn{2}{c}{level-3}& Combined\\
        &~~~$E_T$ (GeV) 
                & Prescale &~~~$E_T$ (GeV) 
                                    & Prescale &~~~$E_T$ (GeV) 
                                                       & Prescale &Prescale\\ \hline
jet20   & 5     & 20,50    & 15     & 12,25    & 20    & 1 & 808     \\
jet50   & 5     & 20,50    & 40     & 1,5      & 50    & 1 & 35      \\
jet70   & 10    & 1,8      & 60     & 8,1      & 70    & 1 & 8       \\
jet100  & 10,20 & 1        & 90     & 1        & 100   & 1 & 1       \\ 
\end{tabular}
\end{ruledtabular}
\label{tab:jet_triggers}
\end{table*}

The jet20, jet50, and jet70 triggers are 
artificially reduced (prescaled)
in order to avoid saturating the
bandwidth of the trigger and data acquisition system.
The jet70 trigger is prescaled by a constant factor of 8 for all data
used in this measurement, 
which means that only one event in eight satisfying the trigger
requirements is accepted.
The prescales for the jet20 and jet50 triggers were changed during
the period this data sample was acquired in order
to accommodate increasing instantaneous luminosity.
The integrated luminosities of the jet50 and jet20 trigger data
samples corrected for the prescale factors are
31.9 and 1.4 $\mbox{pb}^{-1}$, respectively.
The four jet triggers are summarized in Table~\ref{tab:jet_triggers}.
The jet yield distributions as functions of uncorrected jet \pt\
($p_{T}^{\rm CAL}$)~\cite{pT} in the rapidity region
$0.1<|y|<0.7$ before correcting for trigger prescales
are shown in Fig.~\ref{fig:jetpt}.

Cosmic ray and beam-related background events are removed by applying
a cut on missing-$E_T$ significance, $\met/ \sqrt{\sum E_{T}}$~\cite{met}.
The cut threshold varies with the highest-\pt\ jet in the event and is
defined by:
\begin{equation}
  \met/ \sqrt{\sum E_{T}} < \min(3+0.0125 \times p_{T}^{\rm max},6);
\end{equation}
where $p_{T}^{\rm max}$ is the maximum uncorrected jet \pt\ in the event
in units of GeV/{\it c},
and $E_T$ and $\met$ are in units of GeV.
The missing-$E_T$ significance cut is about $100\%$ efficient for
low-\pt\ jets, and the acceptance decreases to about $90\%$ for the
highest-\pt\ jets used in this measurement.

Primary vertices ($p\bar p$ interaction points) are reconstructed from
fits to tracks in each event and from the beam-line constraint, and
the vertex with the highest total $p_T$
of the associated tracks is chosen as the event vertex.
In order to ensure that particles from the $p\bar p$ interactions are
well measured by the CDF II detector,
an event vertex is required to be within 60 cm of the center of
the detector in $z$~\cite{coordinate}.
From the beam profile measured in data,
the acceptance of the event $z$-vertex requirement has been determined
to be $0.958 \pm 0.002$

The inclusive differential jet cross section can be defined as:
\begin{equation}
\frac{ d^2\sigma}{dp_T dy }
= \frac{ 1}{\Delta y}\frac{ 1}{\int {\cal L} dt}\frac{N_{\rm jet} 
}{\Delta p_T },
\end{equation}
%
where $N_{\rm jet}$ is the number of jets in each $p_T$ bin of width
$\Delta p_T$,
$\int {\cal L} dt$ is the effective integrated luminosity which
accounts for trigger prescales, and $\Delta y$ is the rapidity interval.
The number of jets in each $p_T$ bin is counted after jet energies are
corrected on average as described below, and the differential cross
sections are further corrected for the efficiencies of the
$\met/\sqrt{\sum E_{T}}$ and $z$-vertex cuts as well as the bin-to-bin
jet migration effects due to finite energy measurement resolution
as written in Sec.~\ref{sec:unfolding}.
The inclusive differential jet cross section is measured in five jet
rapidity intervals, $|y|<0.1$, $0.1<|y|<0.7$, $0.7<|y|<1.1$,
$1.1<|y|<1.6$, and $1.6<|y|<2.1$ based on detector geometry.

\section{\label{sec:Corrections} Jet Energy and Resolution Corrections}

The jet energies measured by the calorimeters are affected by 
instrumental effects, such as calorimeter non-linearity and energy
smearing due to finite energy resolution of the calorimeters.
These biases are corrected for in several steps as outlined below.
First, an $\eta$-dependent relative correction is applied 
in order to equalize in $\eta$ the response of the calorimeters to
jets. 
The equalized jet \pt\ is then corrected for the pileup effect, \ie\
the effect of additional $p\bar p$ interactions in the same bunch
crossing.
Then, a $p_T$-dependent absolute correction is applied to
correct for the average under-measured hadron energy due to
the non-linear response of the CDF calorimeters.
These corrections are applied on a jet-by-jet basis and
corrected jets are binned in $p_T$.
This binned jet cross section is corrected
for the efficiency of the event selection criteria and bin-to-bin 
jet migration effects due to energy smearing (unfolding).
These corrections are derived by comparing the binned hadron-level
cross section
and the calorimeter-level cross sections corrected by the
aforementioned jet-by-jet
corrections using Monte Carlo events.
After these corrections,
the data have been corrected to the hadron level.
In order to compare data with pQCD predictions, the effects of the
underlying event (UE) and hadronization need to be accounted for,
which is discussed in Sec.~\ref{sec:Theory}.

The Monte Carlo simulation used to derive the corrections, and
the details of each correction step are described below. 

\subsection{Monte Carlo Simulation}
\label{subsec:mc}

The parton shower \MC\ programs \py\ 6.2~\cite{PYTHIA} and \hw\
6.5~\cite{HERWIG} along with the CDF detector simulation
are used to derive the various corrections which are applied to the
data, and to estimate systematic uncertainties in the measurement.
The proton and antiproton PDFs are taken from
\cteq ~\cite{CTEQ5L}.  \py\ Tune A~\cite{TuneA_footnote},
which refers to a set of parameters
chosen to describe observables in the CDF data which are sensitive to
the effects from the underlying event~\cite{UE_CDF,TUNEA}, is used for all
\py\ calculations mentioned in this paper.  Tune A is especially
important for the UE correction discussed in Sec.~\ref{sec:Theory}.

The CDF II detector simulation is based on {\sc geant3}~\cite{GEANT3} in which
a parametrized shower simulation, {\sc gflash}~\cite{GFLASH}, is used to
simulate the energy deposited in the calorimeter.  The {\sc
  gflash} parameters are tuned to test-beam data for electrons and
high-momentum charged pions and to the {\it in-situ} collision data
for electrons from $Z$ decays and low-momentum charged hadrons~\cite{JES}.
The \MC\ simulation is used to derive various jet corrections to be
applied to the data, and to evaluate the associated systematic
uncertainties.  However, the real calorimeter response to jets is not
described perfectly by the calorimeter simulation. 
Differences in the relative jet energy response and jet energy
resolution between the collision data and \MC\ simulation events were
investigated using \pt\ balance in dijet events~\cite{JES}
and the ``bisector'' method~\cite{Bisector}, respectively.

Comparisons of dijet \pt\ balance reveal that the
variation of the jet energy scale with $\eta$ is different for data and
\MC\ and that this difference depends on jet \pt\ at high rapidity
($|y|>1.1$).
For example, the jet energy scale in the plug calorimeter region is
higher in \MC\ than in data by $\sim2\%$ and the difference increases
slightly with jet \pt.
This difference is accounted for by the relative corrections
which are described in detail in Sec.~\ref{subsec:relcor}.

The bisector method allows the jet energy resolution in the
real CDF II detector and in the simulation to be compared.
Events with a dijet topology are used for the study by requiring
that only two jets have $p_T>10$ GeV/{\it c}. 
In order to minimize the effects of pileup, only events with
exactly one reconstructed primary vertex are used.  
Also, one jet is required to be in the central region of the detector
($0.1<|y|<0.7$) and is referred to as the ``trigger'' jet.  The second jet is
called the ``probe'' jet and can be in any other rapidity
region ($|y|<2.1$).  A coordinate system is defined in the transverse plane
with one axis aligned with the bisector of the two jets.
With reference to Fig.~\ref{fig:bisector}, the following
components related to the jet energy resolution are studied as
functions of dijet mean $p_T$:

\begin{itemize}
\item[(a)] $\sigma _{\perp}$ is the R.M.S. of the $\Delta p_{T_{\perp}}$
  distribution where $\Delta p_{T_{\perp}} \equiv
  (p_{T_{1}}+p_{T_{2}})\cos(\Delta \phi
  _{12}/2)$:
  $p_{T_{1}}$, $p_{T_{2}}$, and $\Delta \phi_{12}$ refer to
  the $p_T$ of the leading and next-to-leading jets and the azimuthal
  angle between the leading and next-to-leading jets, respectively.
  This component of the $p_T$ imbalance is more sensitive
  to physics effects.
\item[(b)] $\sigma _{||}$ is the R.M.S. of the $\Delta p_{T_{||}}$
  distribution where $\Delta p_{T_{||}}
  \equiv (p_{T_{1}}-p_{T_{2}})\sin(\Delta \phi
  _{12}/2)$. This
  component of the $p_{T}$ imbalance is sensitive both to physics and
  detector effects.
\item[(c)] $\sigma_{D}$ is the quadratic difference between $\sigma _{||}$
  and $\sigma _{\perp}$ ($\sigma_{D} \equiv
  \sqrt{\sigma_{||}^{2}-\sigma_{\perp}^{2}}$). This should be
  most sensitive to detector effects since the physics effects in
  $\sigma_{\perp}$, which are expected to give an isotropic contribution in the
  transverse plane, are removed.  It should also be noted that since
  both jets are contributing to $\sigma _{D}$, for a single jet
  $\sigma = \sigma_{D} / \sqrt{2}$.
\end{itemize}

For comparing the jet energy resolution in the real CDF II detector to
that in the simulation,
$\sigma_{D}$ is used.
In the rapidity region of $0.1<|y|<0.7$, the detector
simulation reproduces the detector jet energy resolution accurately.
Figure~\ref{fig:bisector_central} shows $\sigma
_{\perp}$, $\sigma _{||}$, and $\sigma_{D}$
for data and \py\ events.
The data to MC ratio of $\sigma_{D}$ is used to compare the jet energy
resolution in the real CDF II detector and in the simulation.  

Figure~\ref{fig:bisector_all} shows the result for the $\sigma_{D}$
ratio in the other rapidity regions. 
In two rapidity regions ($0.7<|y|<1.1$ and $1.6<|y|<2.1$), it was
found that MC slightly underestimates the jet energy resolution in
data; to account for the differences,
extra smearing is applied on jet energies in MC events to match the
jet energy resolution between data and MC when the absolute
and unfolding corrections are derived. The extra jet energy smearing
results in $<6\%$ changes in cross section in most bins, and
$<15\%$ in the most extreme cases. 

\subsection{Jet Corrections}
\subsubsection{Relative Correction}
\label{subsec:relcor}

The calorimeter response to jets is not uniform in $\eta$.
The non-uniformity arises from cracks between calorimeter
modules and also from the different responses of the central and plug
calorimeters.
The relative correction is introduced to make the jet energy response
flat in $\eta$.

The leading two jets in dijet events are expected to balance in \pt\
in the absence of QCD radiation. Therefore, 
\pt\ balance in dijet
events is a useful tool to study the jet energy response as a function of
$\eta$ and to derive the relative correction.
To determine the $\eta$-dependent relative jet energy correction,
a jet with $0.2<|\eta|<0.6$ (where the CDF calorimeter is
well understood) is defined as a trigger jet and the other jet is
defined as
a probe jet.
The $p_T$ balance ($\beta\equiv p_T^{\rm probe}/p_T^{\rm
  trigger}$) of these two jets ~\cite{JES}
as a function of probe-jet $\eta$ is shown in Fig.~\ref{fig:relcor}.
It shows $\beta\sim1$ in the region where
the trigger jet is selected, {\it i.e.}, $0.2<|\eta|<0.6$.
There are dips at $\eta\sim0$ and $\pm1$ which are due to gaps between
the calorimeter modules.

The $\eta$-dependent relative corrections are obtained from 
a fit to the $\beta$ distribution at a given jet $p_T$.
These corrections are derived independently for data and MC.
The $\beta$
ratio for data to MC simulation for all rapidity regions is shown in
Fig.~\ref{fig:djbal} as a function of jet $p_T$.
A small additional \pt-independent correction
is required in the region $|y|<0.1$ to bring MC into agreement
with data.  As mentioned earlier, the data-\MC\ difference in the
relative jet energy scale depends on jet \pt\ at $|y|>1.1$.
Therefore an additional correction with \pt\ dependence is derived
for the two highest $|y|$ regions and is applied in order to match
\MC\ to data at
any jet \pt.  Due to lack of statistics at high \pt\ mainly in data, the
uncertainty associated with this correction increases with increasing
jet \pt\ as indicated by the dashed lines in Fig.~\ref{fig:djbal}.

\subsubsection{Pileup Correction}
\label{subsec:pileup}
Extra $p\bar p$ interactions in the same bunch crossing as the
interaction which produced the jets can contribute energy to the jets.
For the data sample used in this measurement, the average number of
additional $p\bar p$ interactions per event is about two.
The number of reconstructed primary vertices is a good estimator of
the number of interactions in the same bunch crossing.
The correction for the additional $p\bar p$ interactions
is derived by measuring the average \pt\ in a randomly chosen 
cone as a function of the number of primary vertices in a sample of
minimum-bias events triggered only on a CLC coincidence between the
two sides of the detector.
The \pt\ in the randomly chosen cone scales linearly with the number
of additional vertices in the event, and the pileup correction is
derived from the slope of this line.
For each additional vertex reconstructed in the
event, $0.97$~GeV/$c$ is subtracted from the jet $p_T$.

\subsubsection{Absolute Correction}
As particles pass through the CDF calorimeter,
not all of their energy is collected.
The absolute correction is applied to each jet
to compensate for this average energy loss.
The correction is derived by comparing hadron-level and
calorimeter-level jets using \py\ and the CDF detector simulation.
Hadron-level and calorimeter-level jets are matched by their position
in $y-\phi$ space ($\Delta R=\sqrt{(\Delta y)^2 + (\Delta\phi)^2}\le
0.7$).
In Fig.~\ref{fig:avgcor}, the average hadron-level jet \pt\ is
shown as a function of the calorimeter jet \pt\ in each
rapidity region.
These distributions are fit to a fourth-order polynomial and the fit
is applied as a correction to the \pt\ of each jet in the data sample.

\subsection{Unfolding Correction}
\label{sec:unfolding}
The next step in correcting the jet \pt\ distribution to the hadron
level is the unfolding correction, which accounts for smearing effects of
the calorimeter and the efficiency of the event
selection criteria.
The hadron-level and calorimeter-level (after the
jet corrections discussed above have been applied) cross sections 
from the \py\ \MC\ events are compared on a bin-by-bin basis to derive
the unfolding correction.
Since these corrections depend on the jet $p_T$ spectra,
the \py\ events are reweighted to match the jet $p_T$ spectra measured
in data before the correction factors are calculated.
These weights are derived by comparing the data corrected to the
hadron level to
the \py\ prediction.
The unfolding corrections shown in Fig.~\ref{fig:bincor} are
obtained from the weighted \py\ distributions and applied to the data.
The change due to the reweighting of \py\ is small (less than $5\%$)
except in the highest-\pt\ bins where the correction is still less
than $20\%$. 
After the unfolding correction is applied to the data,
the measurement has been corrected for all the instrumental effects
and presented at the hadron level.

\section{\label{sec:Systematics} Systematic Uncertainties}

The systematic uncertainties in the measurement are presented below.

\begin{description}
\item[{\it Jet energy scale.}]
  The uncertainty in the jet energy scale mainly comes from the
  uncertainty in the tuning of the central calorimeter simulation 
  based on the response to individual particles.
  This uncertainty is less than 3\% of the jet energy over the
  entire jet energy range~\cite{JES}.
  The resulting systematic uncertainty in the cross section
  measurement
  varies from $10\%$ at low \pt\ up to $90\%$ at high \pt\ in some
  rapidity regions.
  The fractional uncertainty on the jet cross section in the
  rapidity region $0.1<|y|<0.7$ due to the jet energy scale is
  shown in Fig.~\ref{fig:sys_central}(a).
  
  The jet energy scale uncertainty 
  may be subdivided into a few components with different dependence on
  jet \pt ~\cite{KT_PRD}:
  \begin{enumerate}
  \item[(1)] A $\pm 1.8 \%$ $ p_{T}$-independent component which arises
    from the uncertainty in the
    calorimeter stability in time ($\pm 0.5 \%$), uncertainty 
    in the modeling of the jet fragmentation ($\pm 1.0 \%$),
    uncertainty in the simulation of the electromagnetic calorimeter
    response($\pm 0.5 \%$), and uncertainty in the simulation of the
    calorimeter response at the boundary between calorimeter towers ($\pm
    1.3 \%$)~\cite{JES}.
  \item[(2)] Contributions due to the description of the calorimeter
    response to hadrons in three different momentum ranges~\cite{JES}:
    \begin{enumerate}
    \item[(2.a)] $  p < 12$   GeV/$c$
    \item[(2.b)] $12 <   p < 20$   GeV/$c$
    \item[(2.c)] $  p > 20$   GeV/$c$
    \end{enumerate}
  \end{enumerate}
  These four components [(1), (2.a), (2.b), and (2.c)] are considered
  independent: Each of the four components is considered fully
  correlated in \pt\ and rapidity and is listed in
  Table~\ref{tab:decompose}.
  This decomposition of the jet energy scale uncertainty for the
  region of $0.1<|y|<0.7$ is
  shown as a relative uncertainty on the jet cross section in
  Fig.~\ref{fig:jes_breakdown}. 

\item[{\it Dijet \pt\ balance.}]
  The dependence on the dijet event selection definitions and
  statistical limitations yield a $0.5$\% uncertainty in the relative
  jet energy correction in all rapidity regions.  In addition, at
  high \pt\ there is a $p_T$-dependent uncertainty on the correction in
  the higher rapidity regions ($|y| > 1.1$) due to low statistics. 
  This uncertainty is considered correlated over \pt\ but
  uncorrelated across different rapidity regions.
  The fractional uncertainty on the cross section in
  the region of $0.1<|y|<0.7$ due to this dijet balance systematic 
  uncertainty is shown in Fig.~\ref{fig:sys_central}(c).

\item[{\it Pileup correction.}]
  The pileup correction is obtained from minimum-bias data, and the
  systematic uncertainty is determined so that it covers
  variations from a set of validation measurements.  Measurements of
  the  pileup correction from dijet, photon-jet,
  and $W\to e\nu$ events result in variations of less than $30\%$ and
  this is taken as the  size of the systematic uncertainty.  
  This uncertainty results in less than $3\%$ uncertainty in the
  cross section measurement.  
  This uncertainty is considered fully correlated in \pt\ and
  rapidity.
  The fractional uncertainty on the cross section 
  in the region of $0.1<|y|<0.7$ due to the pileup systematic
  uncertainty is shown in Fig.~\ref{fig:sys_central}(e).

\item[{\it Unfolding and $p_T$-spectra.}]
  The difference between the \py\ and \hw\ predictions is taken as the
  systematic uncertainty on the unfolding correction, as they have
  different fragmentation models.  
  The fractional uncertainty on the cross section 
  in the region of $0.1<|y|<0.7$ due to the jet unfolding systematic
  uncertainty is shown 
  in Fig.~\ref{fig:sys_central}(d).  
  This uncertainty is considered fully correlated in \pt\ and
  rapidity.  
  As mentioned in
  Sec.~\ref{sec:Corrections}, \py\ events are reweighted when the
  unfolding corrections are determined so that the jet
  $p_T$ spectrum agrees with what is observed in data.
  The uncertainty in the unfolding correction due to the dependence
  on the jet $p_T$ spectra is taken conservatively from 
  the change in the unfolding corrections with and without reweighting
  \py\ events. 
  This reweighting is done independently in each rapidity region;
  therefore, the uncertainty is considered correlated over \pt\ but
  uncorrelated across different rapidity regions.
  The fractional uncertainty on the cross section in the
  rapidity region $0.1<|y|<0.7$ due to reweighting is shown in
  Fig.~\ref{fig:sys_central}(f).

\item[{\it Jet energy resolution.}]
  Due to the sharply falling spectrum of the inclusive jet cross
  section, any imperfect modeling of the jet energy smearing in the detector
  simulation will affect the derived unfolding correction.
  The calorimeter-level jets in the \py\ events have been smeared by an 
  extra amount such that $\sigma_D$ as defined in the bisector method
  changes by $10\%$. The effect of this extra smearing on the jet
  differential cross section
  is taken as the systematic uncertainty 
  due to resolution. 
  The jet resolution differences between data and MC events vary
  with rapidity and the corrections are performed 
  independently in five rapidity regions; therefore,
  this uncertainty is considered correlated over \pt\ but
  uncorrelated across different rapidity regions.
  The fractional uncertainty on the cross section in the
  rapidity region $0.1<|y|<0.7$ due to the jet energy resolution is shown in
  Fig.~\ref{fig:sys_central}(b).

\item[{\it Luminosity.}] There is a $6\%$ uncertainty in normalization due
  to the luminosity measurement~\cite{lumi}.
  This uncertainty is considered fully correlated in \pt\ and
  rapidity.

\end{description}

The total systematic uncertainty on the hadron-level jet cross section
for each jet rapidity region is shown in
Fig.~\ref{fig:sys_tot}~\cite{syserr}.
The systematic uncertainties on the measured
cross section from each source for each rapidity region are given in
Tables~\ref{tab:sys1}-\ref{tab:sys5}.

\section{\label{sec:Theory} Theoretical Predictions}

Perturbative QCD calculations for the inclusive jet cross sections
in hadron-hadron collisions have been performed so far only up to
next-to-leading order, and their predictions are provided
at the parton level~\cite{EKS,JETRAD,NLOJET_1,NLOJET_2}
in which the final state is comprised of only two or three partons.
Our measurement is compared with predictions from
{\sc fast}NLO~\cite{FastNLO} which are based on
the {\sc nlojet}++~\cite{NLOJET_1,NLOJET_2} program.
CTEQ6.1M~\cite{CTEQ6.1M} is used for the parton distribution functions
(PDFs).
The renormalization and factorization scales ($\mu_R$ and $\mu_F$)
are chosen to be the transverse momentum of the jet
divided by two,
which is the same as that used in the global QCD
analyses~\cite{CTEQ6.1M,MRST2004} to determine the PDFs.
Using $\mu_R=\mu_F=p_T^{\rm jet}$ 
gives up to $10\%$ smaller predictions in the cross section.
The uncertainties on the predictions due to PDF are estimated by using
the 40 CTEQ6.1M error PDFs~\cite{CTEQ6.1M,Hessian}, and the MRST2004
PDF~\cite{MRST2004} is also used to obtain a prediction.
In order to account for the splitting and merging step of the Midpoint
jet clustering algorithm when clustering partons after the parton
shower or particles after hadronization~\cite{CDF_midpoint2},
a parameter $R_{\rm sep}$~\cite{Rsep} with a value of $1.3$ is used for
the Midpoint algorithm at the NLO parton level.
Two partons are clustered into a single jet if they are within
$R_{\rm cone}$ of the jet centroid and within $R_{\rm cone}\times
R_{\rm sep}$ of
each other.
An $R_{\rm sep}$ value of $2.0$ (\ie\ the Midpoint algorithm without
$R_{\rm sep}$) yields $<5\%$ larger cross sections for NLO pQCD
predictions.

As mentioned earlier, NLO pQCD calculations provide predictions not at
the hadron level, to which the data are corrected, but at the parton
level, {\it i.e.}, they do not account for the underlying event and
hadronization effects.
In order to compare the data corrected to the hadron level with
predictions for jets clustered from partons as obtained from NLO pQCD
calculations, such effects must be accounted for.
The underlying event contributes energy to the jet cone that is not
associated with the hard scattering event, {\it i.e.}, energy from
collisions of other partons in the proton and antiproton.
Hadronization may cause particles originating from partons whose
trajectories lie inside the jet cone to go outside of the jet cone.
The effect of hard gluon emission outside the jet cone
is already accounted for in NLO pQCD predictions, and thus it is not
included in the corrections discussed below. 

The bin-by-bin parton-to-hadron-level ($C_{p\to h}$) corrections 
are obtained by applying the Midpoint clustering algorithm to the
hadron-level and to the parton-level outputs of the
\py\ Tune A dijet Monte Carlo samples, generated with and without an
underlying event. The samples without the underlying event were
generated by turning off multiple parton interactions (MPIs). 
The parton-to-hadron-level correction increases the NLO pQCD cross section
predictions by about 10\% at low $p_T$ and is negligible at high $p_T$
as shown in Fig.~\ref{fig:Par2Had_sys}.  

The uncertainty on the parton-to-hadron-level correction is
estimated from the difference in the predictions for this correction
from \hw\ and \py.  \hw\ does not include
MPIs in its underlying event model, and
instead relies on initial state radiation (ISR) and beam remnants to
populate the underlying event.
The difference between \hw\ and \py\ is conservatively taken
for this systematic uncertainty, and this uncertainty is
represented by the shaded bands in Fig.~\ref{fig:Par2Had_sys}. 

\section{\label{sec:Results} Results}

The measured inclusive differential jet cross sections at the
hadron level are shown in Fig.~\ref{fig:result_dist_relcor_all},
and Tables~\ref{tab:sigma1}-\ref{tab:sigma4} show
the lists of the measured cross sections for each jet \pt\ and
rapidity bin together with the statistical and total
systematic uncertainties, and parton-to-hadron-level correction factors.
The ratios of the measured cross sections to the NLO pQCD predictions
from {\sc fast}NLO (corrected to the hadron level) based on the
CTEQ6.1M PDF are shown in Fig.~\ref{fig:result_rat_relcor_all}
together with the theoretical uncertainties due to PDF.
The measured inclusive jet cross sections tend to be lower but still
in agreement with the NLO pQCD predictions within the experimental and
theoretical uncertainties.

To quantify the comparisons, a procedure based on the $\chi^2$
defined as:
\begin{equation}
  \chi^2 = \sum_{i=1}^{\rm nbin}
  \frac{[\sigma_i^{\rm data}-\sigma_i^{\rm theory}]^2}
  {[(\delta\sigma_{i}^{\rm data-stat})^2 +
    (\delta\sigma_{i}^{\rm theory-stat})^2]}
  + \sum_{j=1}^{\rm nsyst}s_j^2,
\label{eqn:chi2}
\end{equation}
\begin{equation}
  \sigma_{i}^{\rm theory} = \sigma_{i,0}^{\rm theory} 
  + \sum_{j=1}^{\rm nsyst} s_j \times 
  \delta\sigma_{i,j}^{{\rm syst}}
\label{eqn:chi2b}
\end{equation}
is used where $\sigma_i^{\rm data}$ and $\delta\sigma_{i}^{\rm data-stat}$
are the measured cross section and its statistical uncertainty
in the $i$-th data point, and
$\sigma_i^{\rm theory}$ and $\sigma_i^{\rm theory-stat}$ are the corresponding
theoretical prediction and its statistical uncertainty.
The $\sigma_i^{\rm theory}$ may be shifted from the nominal
theoretical prediction for the $i$-th data point, $\sigma_{i,0}^{\rm
  theory}$, as shown in Eq.~(\ref{eqn:chi2b}), where
$\delta\sigma_{i,j}^{\rm syst}$ is the systematic uncertainty in
the $i$-th data point due to the $j$-th systematic uncertainty
and $s_j$ is the standard deviation in the $j$-th systematic uncertainty.
The first sum in Eq.~(\ref{eqn:chi2}) is carried out over all
data points, and the second sum in Eq.~(\ref{eqn:chi2}) and
the sum in Eq.~(\ref{eqn:chi2b}) are over all independent
sources of the systematic uncertainties. 
These systematic shifts $s_j$ are chosen to minimize the $\chi^2$
defined above using the {\sc minuit} program~\cite{MINUIT}.
This $\chi^2$ definition is basically the same as those used in
the previous CDF inclusive jet cross section
measurements~\cite{incjet_run1,KT_PRD}, and this $\chi^2$
is equivalent to the one calculated using the covariance matrix
technique.

In the $\chi^2$ calculation, 
the systematic uncertainties due to jet energy scale (four independent
contributions), luminosity, pileup, and unfolding 
are treated as correlated across all data points in $p_T$ and rapidity.
The uncertainties from dijet $p_T$ balance, 
jet energy resolution, and $p_T$ spectra are treated as
correlated over $p_T$ in a rapidity region but are uncorrelated across
different rapidity regions, as discussed in Sec.~\ref{sec:Systematics}.
As for the theoretical uncertainty, the uncertainty
on $C_{p\to h}$ is considered as fully correlated across all data
points, however the PDF and scale uncertainties are not considered.
This $\chi^2$ test yields the probabilities of
$71$, $91$, $23$, $69$ and $91$\%
when it is performed separately
in the five rapidity regions of
$|y|<0.1$,
$0.1<|y|<0.7$,
$0.7<|y|<1.1$,
$1.1<|y|<1.6$, and
$1.6<|y|<2.1$.
The global $\chi^2$ test which is performed simultaneously on all the
data points in all five rapidity regions yields
the reduced $\chi^2$, $\chi^2/{\rm n.d.f.}=94/72$ corresponding to a
probability of 4\%.

As shown in Fig.~\ref{fig:result_rat_relcor_all},
the experimental uncertainties in the measurement are comparable
or somewhat smaller than the PDF uncertainties on the theoretical
predictions, especially in higher $|y|$ regions, and thus
this measurement will lead to useful constraints on PDFs
when it is included in QCD global fits.

While this measurement was underway, a new cone-based jet
clustering algorithm, called SISCone~\cite{SIS}, was proposed which is
a seedless algorithm and thus infrared safe to all orders in pQCD.
We have studied the impact of using the SISCone algorithm instead of
the Midpoint algorithm in Appendix~\ref{sec:SISCone} and found that
the ratio of the measured cross section over theoretical predictions
would change by only $\sim$1\%. 
Therefore, both algorithms will yield similar data-theory
comparisons and lead to a similar PDF parametrization when the
measurement is included in QCD global fits.

\section{\label{sec:CompareKt} 
  Comparison with the Measurement using $k_T$ Clustering Algorithm}

As mentioned in Sec.~\ref{sec:Intro}, the CDF collaboration has 
recently
made a measurement of the inclusive jet cross section using
the $k_T$ jet clustering algorithm~\cite{KT_PRD}.
In this section, our measurement is compared with the results
obtained with the $k_T$ algorithm with $D=0.7$ by taking the
ratio of the cross
sections from the two measurements and comparing it with theoretical
predictions.
In order to make a useful comparison,
the correlations between the statistical and systematic uncertainties
were studied and are presented below.

\subsection{Statistical Correlation}
The datasets used in the two measurements have about 90\%
overlap, and even in the same events the Midpoint and $k_T$
algorithms may lead to a different set of jets and thus populate different
$p_T$ bins which are treated as statistically independent.
In order to study the statistical correlation between the two
measurements,
both $k_T$ and Midpoint jet clustering algorithms are applied
in events used in both measurements and if the resulting jets from
both algorithms are matched in $y-\phi$ space within $R<0.7$ and
fall into the same jet $p_T$ and rapidity bin, those jets are treated
as correlated, 
otherwise they are considered uncorrelated.
This was done for the data points for which events from the
jet100 and jet70 triggers are used. The situation is more complicated
for data points from jet20 and jet50 triggers where prescale factors were
changed during the data taking period.
Thus, the statistical uncertainties are treated
as uncorrelated in the two measurements for data points for which the
jet20 and jet50 trigger events are used.
It should be noted
that the statistics are high in these triggers and 
the statistical uncertainties are small compared with the systematic
uncertainties.

\subsection{Systematic Correlation}
The systematic uncertainties arising from the jet energy scale, 
unfolding correction, and underlying event modeling
were determined with the same methods in the two measurements, and
thus these uncertainties are treated as fully correlated between the
two measurements, {\it i.e.}, the systematic uncertainties are
canceled in the ratio of the two measurements.

In this analysis, the pileup correction is determined 
by measuring the average \pt\ in a randomly chosen 
cone as a function of the number of primary vertices in minimum-bias
data as discussed in Sec.~\ref{sec:Corrections}. However, a different
method is used in the measurement using the $k_T$
algorithm~\cite{KT_PRD}, and thus
uncertainties arising from pileup corrections are treated as
uncorrelated.
The details of $\eta$-dependent jet corrections are also compared
and it is concluded that this uncertainty is also uncorrelated. 

The correction for the jet $p_T$ resolution and the associated
systematic uncertainties between the two measurements are determined
in a similar way in the two measurements but 
the size of the correction was found to be different. 
The jet $p_T$ resolution difference between data and MC events is
measured in dijet events with third-jet $p_T<$10 GeV/$c$ in order to
use dijet events with a clear back-to-back structure,
and this dijet event selection is not equivalent when jets are
clustered by the $k_T$ algorithm and by the Midpoint algorithm.
In addition, the jet resolution correction is sensitive to the procedure
of applying the $\eta$-dependent relative jet correction.
By varying the dijet selection requirement and the $\eta$-dependent
jet correction procedure,
35\% of the size of the jet $p_T$ resolution uncertainty
is found to be uncorrelated between the two measurements.

\subsection{Results}
The ratio of the cross section measured with the $k_{T}$ algorithm
to that with the Midpoint algorithm is shown
in Fig.~\ref{fig:kt_mp_ratio_all}.
This ratio of the NLO pQCD predictions as given
by {\sc fast}NLO (corrected to the hadron level) and the ratio from
\py\ are also included.
It should be noted that the rapidity region where the agreement is
only marginal ($0.7<|y|<1.1$) corresponds to the crack between the 
central and plug calorimeters.
In the other regions, good agreement is observed over a large range of
rapidity and \pt.  This agreement means that both algorithms observe
similar systematic trends when compared to NLO pQCD predictions
and favor the same PDF parametrization.  
In addition, the agreement between the data, \py, and NLO pQCD predictions for
these ratios provide strong evidence that these clustering algorithms
are behaving in a consistent way when clustering particles at the
parton, hadron, and calorimeter-tower (detector) levels.

\section{\label{sec:Conclusions} Conclusions}

A measurement has been presented of the inclusive jet cross section for jets
clustered by the Midpoint jet-finding algorithm using 1.13
fb$^{-1}$ of data collected by the CDF experiment.
The measured cross sections tend to be lower than the central NLO pQCD
predictions, but they are still consistent when systematic
uncertainties are taken into account.
Similar trends are also observed in the recent results from CDF using
the $k_T$ algorithm~\cite{KT_PRD} and from D0 using the Midpoint
algorithm~\cite{Dzero_Run2_incjet}.
In the forward regions, 
the measurement precision is better than current PDF uncertainties.
When included in QCD global fits 
this will provide further constraints on PDFs, especially the gluon
distributions at high $x$.
Since the measured cross sections tend to be lower than the central
NLO pQCD predictions, the inclusion of this measurement to QCD global
fits will lead to somewhat reduced gluon densities at high $x$.
The results are also compared to the recent 
measurement of the inclusive jet cross section using the $k_{T}$ jet
clustering algorithm~\cite{KT_PRD}, and
it is found that the ratios of the cross sections measured with the two
algorithms are in reasonable agreement with theoretical expectations.

\begin{acknowledgments}
We thank the Fermilab staff and the technical staffs of the
participating institutions for their vital contributions. This work
was supported by the U.S. Department of Energy and National Science
Foundation; the Italian Istituto Nazionale di Fisica Nucleare; the
Ministry of Education, Culture, Sports, Science and Technology of
Japan; the Natural Sciences and Engineering Research Council of
Canada; the National Science Council of the Republic of China; the
Swiss National Science Foundation; the A.P. Sloan Foundation; the
Bundesministerium f\"ur Bildung und Forschung, Germany; the Korean
Science and Engineering Foundation and the Korean Research Foundation;
the Science and Technology Facilities Council and the Royal Society,
UK; the Institut National de Physique Nucleaire et Physique des
Particules/CNRS; the Russian Foundation for Basic Research; the
Ministerio de Educaci\'{o}n y Ciencia and Programa Consolider-Ingenio
2010, Spain; the Slovak R\&D Agency; and the Academy of Finland.
\end{acknowledgments}

\appendix
\section{Seedless Infrared-Safe Cone Algorithm}
\label{sec:SISCone}

Recently, a cone algorithm (SISCone) has been proposed
which is a seedless algorithm and thus infrared safe to all orders in
pQCD~\cite{SIS}.
One of the main problems with the use of a seedless cone algorithm has been
its slow speed with respect to the seeded cone algorithms (such as the
Midpoint algorithm); however,
the SISCone algorithm has a speed comparable to the seeded cone
algorithms.
We have studied the differences in the inclusive jet cross section between
the Midpoint algorithm used in this paper and the SISCone algorithm
using \py\ Monte Carlo samples.
Studies with the \py\ samples generated with the Tune A parameters show
that, at the hadron level, the SISCone algorithm yields
the inclusive jet cross section lower than the Midpoint algorithm
by $\sim5$\% at low \pt\ and $\sim2$\% in the highest \pt\ bins 
independent of jet rapidities; however, the
\py\ samples generated without multiple parton interactions show that
the parton-level inclusive jet cross section is consistent between
the Midpoint algorithm and SISCone algorithm to better than 1\%,
if the same cone radius and the same merging fraction $f_{\rm merge}$ are
used for both algorithms~\cite{JetReview}.
Therefore, although the inclusive jet cross section measured
at the hadron level will decrease by up to $\sim5$\% with the SISCone
algorithm, the change will be compensated by the
parton-to-hadron-level corrections applied to the NLO pQCD
predictions, and thus,
the comparisons between the measured cross section and NLO pQCD
predictions will essentially be the same.

\bibliography{MP_PRD}

\begin{thebibliography}{57}
\expandafter\ifx\csname natexlab\endcsname\relax\def\natexlab#1{#1}\fi
\expandafter\ifx\csname bibnamefont\endcsname\relax
  \def\bibnamefont#1{#1}\fi
\expandafter\ifx\csname bibfnamefont\endcsname\relax
  \def\bibfnamefont#1{#1}\fi
\expandafter\ifx\csname citenamefont\endcsname\relax
  \def\citenamefont#1{#1}\fi
\expandafter\ifx\csname url\endcsname\relax
  \def\url#1{\texttt{#1}}\fi
\expandafter\ifx\csname urlprefix\endcsname\relax\def\urlprefix{URL }\fi
\providecommand{\bibinfo}[2]{#2}
\providecommand{\eprint}[2][]{\url{#2}}

\bibitem[{err()}]{erratum}
\bibinfo{note}{Figures~\ref{fig:sys_tot},
  \ref{fig:result_dist_relcor_all}--\ref{fig:kt_mp_ratio_all} and
  Tables~\ref{tab:sys4} and \ref{tab:sigma1}--\ref{tab:sigma4} were corrected
  in May, 2009. In prior versions, Table~\ref{tab:sys4} had the columns for the
  systematic uncertainties from the unfolding correction, $p_T$-spectra
  modeling, and jet energy resolution modeling mislabeled, and the final set of
  unfolding corrections did not propagate into
  Figs.~\ref{fig:result_dist_relcor_all}--\ref{fig:kt_mp_ratio_all} and
  Tables~\ref{tab:sigma1}--\ref{tab:sigma4}. Miscalculations in error
  propagation were also corrected which affected Figs.~\ref{fig:sys_tot},
  \ref{fig:result_dist_relcor_all}--\ref{fig:kt_mp_ratio_all} and
  Tables~\ref{tab:sigma1}--\ref{tab:sigma4}. It should be noted that the
  comparisons of data and theoretical predictions presented in
  Sec.~\ref{sec:Results} used the correct numbers and thus the conclusions of
  the paper were not affected by these corrections.}

\bibitem[{\citenamefont{Eichten \emph{et~al.}}(1983)\citenamefont{Eichten,
  Lane, and Peskin}}]{NewPhys1}
\bibinfo{author}{\bibfnamefont{E.~J.} \bibnamefont{Eichten}},
  \bibinfo{author}{\bibfnamefont{K.~D.} \bibnamefont{Lane}}, \bibnamefont{and}
  \bibinfo{author}{\bibfnamefont{M.~E.} \bibnamefont{Peskin}},
  \bibinfo{journal}{Phys. Rev. Lett.} \textbf{\bibinfo{volume}{50}},
  \bibinfo{pages}{811} (\bibinfo{year}{1983}).

\bibitem[{New()}]{NewPhys2}
\bibinfo{note}{K. Lane, arXiv:hep-ph/9605257.}

\bibitem[{\citenamefont{Gross and Wilczek}(1973)}]{pqcd1}
\bibinfo{author}{\bibfnamefont{D.~J.} \bibnamefont{Gross}} \bibnamefont{and}
  \bibinfo{author}{\bibfnamefont{F.}~\bibnamefont{Wilczek}},
  \bibinfo{journal}{Phys. Rev. D} \textbf{\bibinfo{volume}{8}},
  \bibinfo{pages}{3633} (\bibinfo{year}{1973}).

\bibitem[{\citenamefont{Fritzsch \emph{et~al.}}(1973)\citenamefont{Fritzsch,
  Gell-Mann, and Leutwyler}}]{pqcd2}
\bibinfo{author}{\bibfnamefont{H.}~\bibnamefont{Fritzsch}},
  \bibinfo{author}{\bibfnamefont{M.}~\bibnamefont{Gell-Mann}},
  \bibnamefont{and}
  \bibinfo{author}{\bibfnamefont{H.}~\bibnamefont{Leutwyler}},
  \bibinfo{journal}{Phys. Lett. B} \textbf{\bibinfo{volume}{47}},
  \bibinfo{pages}{365} (\bibinfo{year}{1973}).

\bibitem[{qcd()}]{qcdbook}
\bibinfo{note}{R.~K.~Ellis, W.~J.~Stirling and B.~R.~Webber, Camb.\ Monogr.\
  Part.\ Phys.\ Nucl.\ Phys.\ Cosmol.\ {\bf 8}, 1 (1996).}

\bibitem[{\citenamefont{Abulencia
  \emph{et~al.}}(2006{\natexlab{a}})}]{Midpoint_RC}
\bibinfo{author}{\bibfnamefont{A.}~\bibnamefont{Abulencia}}
  \bibnamefont{\emph{et~al.}} (\bibinfo{collaboration}{CDF Collaboration}),
  \bibinfo{journal}{Phys. Rev. D} \textbf{\bibinfo{volume}{74}},
  \bibinfo{pages}{071103} (\bibinfo{year}{2006}{\natexlab{a}}).

\bibitem[{\citenamefont{Abulencia \emph{et~al.}}(2006{\natexlab{b}})}]{KT_PRL}
\bibinfo{author}{\bibfnamefont{A.}~\bibnamefont{Abulencia}}
  \bibnamefont{\emph{et~al.}} (\bibinfo{collaboration}{CDF Collaboration}),
  \bibinfo{journal}{Phys. Rev. Lett.} \textbf{\bibinfo{volume}{96}},
  \bibinfo{pages}{122001} (\bibinfo{year}{2006}{\natexlab{b}}).

\bibitem[{\citenamefont{Abulencia \emph{et~al.}}(2007)}]{KT_PRD}
\bibinfo{author}{\bibfnamefont{A.}~\bibnamefont{Abulencia}}
  \bibnamefont{\emph{et~al.}} (\bibinfo{collaboration}{CDF Collaboration}),
  \bibinfo{journal}{Phys. Rev. D} \textbf{\bibinfo{volume}{75}},
  \bibinfo{pages}{092006} (\bibinfo{year}{2007}).

\bibitem[{\citenamefont{Abazov \emph{et~al.}}(2008)}]{Dzero_Run2_incjet}
\bibinfo{author}{\bibfnamefont{V.~M.} \bibnamefont{Abazov}}
  \bibnamefont{\emph{et~al.}} (\bibinfo{collaboration}{D0 Collaboration}),
  \bibinfo{journal}{Phys. Rev. Lett.} \textbf{\bibinfo{volume}{101}},
  \bibinfo{pages}{062001} (\bibinfo{year}{2008}).

\bibitem[{coo()}]{coordinate}
\bibinfo{note}{We use a cylindrical coordinate system with the $z$ coordinate
  along the proton beam direction and the origin at the center of the detector,
  the azimuthal angle $\phi$, and the polar angle $\theta$ usually expressed
  through the pseudorapidity $\eta=-\ln\tan(\theta/2)$. The rapidity $y$ is
  defined as $y=1/2\ln((E+p_z)/(E-p_z))$ where $E$ denotes the energy and $p_z$
  is the momentum component along $z$. The transverse energy and transverse
  momentum are given by $E_{T}=E\sin(\theta)$ and $p_{T}=p\sin(\theta)$ where
  $p$ is the magnitude of the momentum vector.}

\bibitem[{CTE()}]{CTEQ6.1M}
\bibinfo{note}{D. Stump \emph{et~al.}, J. High Energy Phys. 10 (2003) 046.}

\bibitem[{CDF({\natexlab{a}})}]{CDF_midpoint1}
\bibinfo{note}{G. C. Blazey {\it et al.}, arXiv:hep-ex/0005012.}

\bibitem[{\citenamefont{Ellis \emph{et~al.}}(2008)}]{JetReview}
\bibinfo{author}{\bibfnamefont{S.~D.} \bibnamefont{Ellis}}
  \bibnamefont{\emph{et~al.}}, \bibinfo{journal}{Prog. Part. Nucl. Phys.}
  \textbf{\bibinfo{volume}{60}}, \bibinfo{pages}{484} (\bibinfo{year}{2008}).

\bibitem[{Les()}]{LesHouches_tool}
\bibinfo{note}{C.~Buttar {\it et al.}, arXiv:0803.0678.}

\bibitem[{TeV()}]{TeV4LHC}
\bibinfo{note}{M. G. Albrow {\it et al.}, arXiv:hep-ph/0610012.}

\bibitem[{sea()}]{search_cone}
\bibinfo{note}{In the previous CDF measurement~\cite{Midpoint_RC}, an
  additional step called the ``Search Cone'' was included in the Midpoint
  algorithm. This step is removed in this measurement as recommended by the
  TeV4LHC QCD Working Group. For further discussions, see Ref.~\cite{TeV4LHC}.}

\bibitem[{\citenamefont{Martin \emph{et~al.}}(2004)\citenamefont{Martin,
  Roberts, Stirling, and Thorne}}]{MRST2004}
\bibinfo{author}{\bibfnamefont{A.~D.} \bibnamefont{Martin}},
  \bibinfo{author}{\bibfnamefont{R.~G.} \bibnamefont{Roberts}},
  \bibinfo{author}{\bibfnamefont{W.~J.} \bibnamefont{Stirling}},
  \bibnamefont{and} \bibinfo{author}{\bibfnamefont{R.~S.}
  \bibnamefont{Thorne}}, \bibinfo{journal}{Phys. Lett. B}
  \textbf{\bibinfo{volume}{604}}, \bibinfo{pages}{61} (\bibinfo{year}{2004}).

\bibitem[{\citenamefont{Ellis and Soper}(1993{\natexlab{a}})}]{KTalgo_Ellis}
\bibinfo{author}{\bibfnamefont{S.~D.} \bibnamefont{Ellis}} \bibnamefont{and}
  \bibinfo{author}{\bibfnamefont{D.~E.} \bibnamefont{Soper}},
  \bibinfo{journal}{Phys. Rev. D} \textbf{\bibinfo{volume}{48}},
  \bibinfo{pages}{3160} (\bibinfo{year}{1993}{\natexlab{a}}).

\bibitem[{\citenamefont{Abbott \emph{et~al.}}(1999)}]{incjet_run1_dzero}
\bibinfo{author}{\bibfnamefont{B.}~\bibnamefont{Abbott}}
  \bibnamefont{\emph{et~al.}} (\bibinfo{collaboration}{D0 Collaboration}),
  \bibinfo{journal}{Phys. Rev. Lett.} \textbf{\bibinfo{volume}{82}},
  \bibinfo{pages}{2451} (\bibinfo{year}{1999}).

\bibitem[{\citenamefont{Abbott \emph{et~al.}}(2002)}]{incjet_run1_kt_dzero}
\bibinfo{author}{\bibfnamefont{B.}~\bibnamefont{Abbott}}
  \bibnamefont{\emph{et~al.}} (\bibinfo{collaboration}{D0 Collaboration}),
  \bibinfo{journal}{Phys. Lett. B} \textbf{\bibinfo{volume}{525}},
  \bibinfo{pages}{221} (\bibinfo{year}{2002}).

\bibitem[{\citenamefont{Acosta \emph{et~al.}}(2005)}]{CDF_detect3}
\bibinfo{author}{\bibfnamefont{D.}~\bibnamefont{Acosta}}
  \bibnamefont{\emph{et~al.}} (\bibinfo{collaboration}{CDF Collaboration}),
  \bibinfo{journal}{Phys. Rev. D} \textbf{\bibinfo{volume}{71}},
  \bibinfo{pages}{052003} (\bibinfo{year}{2005}).

\bibitem[{\citenamefont{Sill}(2000)}]{SVXII}
\bibinfo{author}{\bibfnamefont{A.}~\bibnamefont{Sill}}, \bibinfo{journal}{Nucl.
  Instrum. Methods Phys. Res., Sect. A} \textbf{\bibinfo{volume}{447}},
  \bibinfo{pages}{1} (\bibinfo{year}{2000}).

\bibitem[{\citenamefont{Affolder \emph{et~al.}}(2000)}]{ISL}
\bibinfo{author}{\bibfnamefont{T.}~\bibnamefont{Affolder}}
  \bibnamefont{\emph{et~al.}}, \bibinfo{journal}{Nucl. Instrum. Methods Phys.
  Res., Sect. A} \textbf{\bibinfo{volume}{453}}, \bibinfo{pages}{84}
  (\bibinfo{year}{2000}).

\bibitem[{\citenamefont{Affolder \emph{et~al.}}(2004)}]{COT}
\bibinfo{author}{\bibfnamefont{T.}~\bibnamefont{Affolder}}
  \bibnamefont{\emph{et~al.}}, \bibinfo{journal}{Nucl. Instrum. Methods Phys.
  Res., Sect. A} \textbf{\bibinfo{volume}{526}}, \bibinfo{pages}{249}
  (\bibinfo{year}{2004}).

\bibitem[{\citenamefont{Balka \emph{et~al.}}(1988)}]{cem}
\bibinfo{author}{\bibfnamefont{L.}~\bibnamefont{Balka}}
  \bibnamefont{\emph{et~al.}}, \bibinfo{journal}{Nucl. Instrum. Methods Phys.
  Res., Sect. A} \textbf{\bibinfo{volume}{267}}, \bibinfo{pages}{272}
  (\bibinfo{year}{1988}).

\bibitem[{\citenamefont{Hahn \emph{et~al.}}(1988)}]{cem_calib}
\bibinfo{author}{\bibfnamefont{S.~R.} \bibnamefont{Hahn}}
  \bibnamefont{\emph{et~al.}}, \bibinfo{journal}{Nucl. Instr. Methods Phys.
  Res. A} \textbf{\bibinfo{volume}{267}}, \bibinfo{pages}{351}
  (\bibinfo{year}{1988}).

\bibitem[{\citenamefont{Bertolucci \emph{et~al.}}(1988)}]{cha_wha}
\bibinfo{author}{\bibfnamefont{S.}~\bibnamefont{Bertolucci}}
  \bibnamefont{\emph{et~al.}}, \bibinfo{journal}{Nucl. Instrum. Methods Phys.
  Res., Sect. A} \textbf{\bibinfo{volume}{267}}, \bibinfo{pages}{301}
  (\bibinfo{year}{1988}).

\bibitem[{\citenamefont{Oishi}(2000)}]{pcal}
\bibinfo{author}{\bibfnamefont{R.}~\bibnamefont{Oishi}},
  \bibinfo{journal}{Nucl. Instrum. Methods Phys. Res., Sect. A}
  \textbf{\bibinfo{volume}{453}}, \bibinfo{pages}{227} (\bibinfo{year}{2000}).

\bibitem[{\citenamefont{Albrow \emph{et~al.}}(2002)}]{pem_testbeam}
\bibinfo{author}{\bibfnamefont{M.}~\bibnamefont{Albrow}}
  \bibnamefont{\emph{et~al.}}, \bibinfo{journal}{Nucl. Instrum. Methods Phys.
  Res., Sect. A} \textbf{\bibinfo{volume}{480}}, \bibinfo{pages}{524}
  (\bibinfo{year}{2002}).

\bibitem[{\citenamefont{Acosta \emph{et~al.}}(2002)}]{lumi}
\bibinfo{author}{\bibfnamefont{D.}~\bibnamefont{Acosta}}
  \bibnamefont{\emph{et~al.}}, \bibinfo{journal}{Nucl. Instrum. Methods Phys.
  Res., Sect. A} \textbf{\bibinfo{volume}{494}}, \bibinfo{pages}{57}
  (\bibinfo{year}{2002}).

\bibitem[{\citenamefont{Affolder \emph{et~al.}}(2001)}]{incjet_run1}
\bibinfo{author}{\bibfnamefont{T.}~\bibnamefont{Affolder}}
  \bibnamefont{\emph{et~al.}} (\bibinfo{collaboration}{CDF Collaboration}),
  \bibinfo{journal}{Phys. Rev. D} \textbf{\bibinfo{volume}{64}},
  \bibinfo{pages}{032001} (\bibinfo{year}{2001}).

\bibitem[{\citenamefont{Kilgore and Giele}(1997)}]{Giele_Kilgore}
\bibinfo{author}{\bibfnamefont{W.~B.} \bibnamefont{Kilgore}} \bibnamefont{and}
  \bibinfo{author}{\bibfnamefont{W.~T.} \bibnamefont{Giele}},
  \bibinfo{journal}{Phys. Rev.} \textbf{\bibinfo{volume}{D55}},
  \bibinfo{pages}{7183} (\bibinfo{year}{1997}).

\bibitem[{had()}]{hadron_level}
\bibinfo{note}{The hadron level in the Monte Carlo generators is defined using
  all final state particles with lifetime above $10^{-11} {\rm s}$.}

\bibitem[{pT()}]{pT}
\bibinfo{note}{$p_{T}^{\rm JET}$ refers to the transverse momentum of jets
  after the absolute jet energy correction has been made while $p_{T}^{\rm
  CAL}$ refers to uncorrected jet $p_{T}$.}

\bibitem[{met()}]{met}
\bibinfo{note}{The missing transverse energy ($\met$) is defined by $\met =
  |\mathbin{\vec{E}\mkern - 11mu/_T}|$, $\mathbin{\vec{E}\mkern - 11mu/_T} =
  -\sum_{i} E_T^i {\bf \hat{n}_i}$, where ${\bf \hat{n}_i}$ is a unit vector
  perpendicular to the beam axis and pointing at the i$^{th}$ calorimeter
  tower. The sum is over all the calorimeter towers with $E_T>100$ MeV and
  $|\eta| < 3.6$. We define the missing $E_T$ significance as $\met/ \sqrt{\sum
  E_{T}} = |\mathbin{\vec{E}\mkern - 11mu/_T}|/\sqrt{\sum_{i} E_T^i}$.}

\bibitem[{\citenamefont{Sjostrand \emph{et~al.}}(2001)}]{PYTHIA}
\bibinfo{author}{\bibfnamefont{T.}~\bibnamefont{Sjostrand}}
  \bibnamefont{\emph{et~al.}}, \bibinfo{journal}{Comput. Phys. Commun.}
  \textbf{\bibinfo{volume}{135}}, \bibinfo{pages}{238} (\bibinfo{year}{2001}).

\bibitem[{HER()}]{HERWIG}
\bibinfo{note}{G. Corcella \emph{et~al.}, J. High Energy Phys. 01 (2001) 010.}

\bibitem[{\citenamefont{Lai \emph{et~al.}}(2000)}]{CTEQ5L}
\bibinfo{author}{\bibfnamefont{H.~L.} \bibnamefont{Lai}}
  \bibnamefont{\emph{et~al.}} (\bibinfo{collaboration}{CTEQ Collaboration}),
  \bibinfo{journal}{Eur. Phys. J. C} \textbf{\bibinfo{volume}{12}},
  \bibinfo{pages}{375} (\bibinfo{year}{2000}).

\bibitem[{Tun()}]{TuneA_footnote}
\bibinfo{note}{R. Field, presented at {\em Fermilab ME/MC Tuning Workshop},
  Fermilab, October 4, 2002. \py\ Tune A implies that the following parameters
  are set in \py\ with \cteq: PARP(67)=4, MSTP(82)=4, PARP(82)=2, PARP(84)=0.4,
  PARP(85)=0.9, PARP(86)=0.95, PARP(89)=1800, PARP(90)=0.25.}

\bibitem[{\citenamefont{Affolder \emph{et~al.}}(2002)}]{UE_CDF}
\bibinfo{author}{\bibfnamefont{T.}~\bibnamefont{Affolder}}
  \bibnamefont{\emph{et~al.}} (\bibinfo{collaboration}{CDF Collaboration}),
  \bibinfo{journal}{Phys. Rev. D} \textbf{\bibinfo{volume}{65}},
  \bibinfo{pages}{092002} (\bibinfo{year}{2002}).

\bibitem[{TUN()}]{TUNEA}
\bibinfo{note}{R. Field and R. C. Group (CDF Collaboration),
  arXiv:hep-ph/0510198.}

\bibitem[{\citenamefont{Brun \emph{et~al.}}(1987)\citenamefont{Brun, Bruyant,
  Maire, McPherson, and Zanarini}}]{GEANT3}
\bibinfo{author}{\bibfnamefont{R.}~\bibnamefont{Brun}},
  \bibinfo{author}{\bibfnamefont{F.}~\bibnamefont{Bruyant}},
  \bibinfo{author}{\bibfnamefont{M.}~\bibnamefont{Maire}},
  \bibinfo{author}{\bibfnamefont{A.~C.} \bibnamefont{McPherson}},
  \bibnamefont{and} \bibinfo{author}{\bibfnamefont{P.}~\bibnamefont{Zanarini}}
  (\bibinfo{year}{1987}), \bibinfo{note}{cern-DD/EE/84-1}.

\bibitem[{\citenamefont{Grindhammer
  \emph{et~al.}}(1990)\citenamefont{Grindhammer, Rudowicz, and
  Peters}}]{GFLASH}
\bibinfo{author}{\bibfnamefont{G.}~\bibnamefont{Grindhammer}},
  \bibinfo{author}{\bibfnamefont{M.}~\bibnamefont{Rudowicz}}, \bibnamefont{and}
  \bibinfo{author}{\bibfnamefont{S.}~\bibnamefont{Peters}},
  \bibinfo{journal}{Nucl. Instrum. Methods Phys. Res., Sect. A}
  \textbf{\bibinfo{volume}{290}}, \bibinfo{pages}{469} (\bibinfo{year}{1990}).

\bibitem[{\citenamefont{Bhatti \emph{et~al.}}(2006)}]{JES}
\bibinfo{author}{\bibfnamefont{A.}~\bibnamefont{Bhatti}}
  \bibnamefont{\emph{et~al.}}, \bibinfo{journal}{Nucl. Instrum. Methods Phys.
  Res., Sect. A} \textbf{\bibinfo{volume}{566}}, \bibinfo{pages}{375}
  (\bibinfo{year}{2006}).

\bibitem[{\citenamefont{Bagnaia \emph{et~al.}}(1984)}]{Bisector}
\bibinfo{author}{\bibfnamefont{P.}~\bibnamefont{Bagnaia}}
  \bibnamefont{\emph{et~al.}} (\bibinfo{collaboration}{UA2 Collaboration}),
  \bibinfo{journal}{Phys. Lett. B} \textbf{\bibinfo{volume}{144}},
  \bibinfo{pages}{283} (\bibinfo{year}{1984}).

\bibitem[{sys()}]{syserr}
\bibinfo{note}{The 6\% uncertainty on the luminosity is not included.}

\bibitem[{\citenamefont{Ellis \emph{et~al.}}(1990)\citenamefont{Ellis, Kunszt,
  and Soper}}]{EKS}
\bibinfo{author}{\bibfnamefont{S.~D.} \bibnamefont{Ellis}},
  \bibinfo{author}{\bibfnamefont{Z.}~\bibnamefont{Kunszt}}, \bibnamefont{and}
  \bibinfo{author}{\bibfnamefont{D.~E.} \bibnamefont{Soper}},
  \bibinfo{journal}{Phys. Rev. Lett.} \textbf{\bibinfo{volume}{64}},
  \bibinfo{pages}{2121} (\bibinfo{year}{1990}).

\bibitem[{\citenamefont{Giele \emph{et~al.}}(1993)\citenamefont{Giele, Glover,
  and Kosower}}]{JETRAD}
\bibinfo{author}{\bibfnamefont{W.~T.} \bibnamefont{Giele}},
  \bibinfo{author}{\bibfnamefont{E.~W.~N.} \bibnamefont{Glover}},
  \bibnamefont{and} \bibinfo{author}{\bibfnamefont{D.~A.}
  \bibnamefont{Kosower}}, \bibinfo{journal}{Nucl. Phys.}
  \textbf{\bibinfo{volume}{B403}}, \bibinfo{pages}{633} (\bibinfo{year}{1993}).

\bibitem[{\citenamefont{Nagy}(2003)}]{NLOJET_1}
\bibinfo{author}{\bibfnamefont{Z.}~\bibnamefont{Nagy}}, \bibinfo{journal}{Phys.
  Rev. D} \textbf{\bibinfo{volume}{68}}, \bibinfo{pages}{094002}
  (\bibinfo{year}{2003}).

\bibitem[{\citenamefont{Catani and Seymour}(1997)}]{NLOJET_2}
\bibinfo{author}{\bibfnamefont{S.}~\bibnamefont{Catani}} \bibnamefont{and}
  \bibinfo{author}{\bibfnamefont{M.~H.} \bibnamefont{Seymour}},
  \bibinfo{journal}{Nucl. Phys.} \textbf{\bibinfo{volume}{B485}},
  \bibinfo{pages}{291} (\bibinfo{year}{1997}).

\bibitem[{Fas()}]{FastNLO}
\bibinfo{note}{T. Kluge, K. Rabbertz, and M. Wobisch, arXiv:hep-ph/0609285.}

\bibitem[{\citenamefont{Pumplin \emph{et~al.}}(2001)}]{Hessian}
\bibinfo{author}{\bibfnamefont{J.}~\bibnamefont{Pumplin}}
  \bibnamefont{\emph{et~al.}}, \bibinfo{journal}{Phys. Rev. D}
  \textbf{\bibinfo{volume}{65}}, \bibinfo{pages}{014013}
  (\bibinfo{year}{2001}).

\bibitem[{CDF({\natexlab{b}})}]{CDF_midpoint2}
\bibinfo{note}{S. D. Ellis, J. Huston, and M. Tonnesmann, eConf C010630, P513
  (2001).}

\bibitem[{\citenamefont{Ellis and Soper}(1993{\natexlab{b}})}]{Rsep}
\bibinfo{author}{\bibfnamefont{S.~D.} \bibnamefont{Ellis}} \bibnamefont{and}
  \bibinfo{author}{\bibfnamefont{D.~E.} \bibnamefont{Soper}},
  \bibinfo{journal}{Phys. Rev. D} \textbf{\bibinfo{volume}{48}},
  \bibinfo{pages}{3160} (\bibinfo{year}{1993}{\natexlab{b}}).

\bibitem[{\citenamefont{James and Roos}(1975)}]{MINUIT}
\bibinfo{author}{\bibfnamefont{F.}~\bibnamefont{James}} \bibnamefont{and}
  \bibinfo{author}{\bibfnamefont{M.}~\bibnamefont{Roos}},
  \bibinfo{journal}{Comput. Phys. Commun.} \textbf{\bibinfo{volume}{10}},
  \bibinfo{pages}{343} (\bibinfo{year}{1975}).

\bibitem[{SIS()}]{SIS}
\bibinfo{note}{G. P. Salam and G. Soyez, J. High Energy Phys. 05 (2007) 086.}

\end{thebibliography}


\begin{widetext}

\begin{figure}[p]
\includegraphics[width=\dfigurewidth]{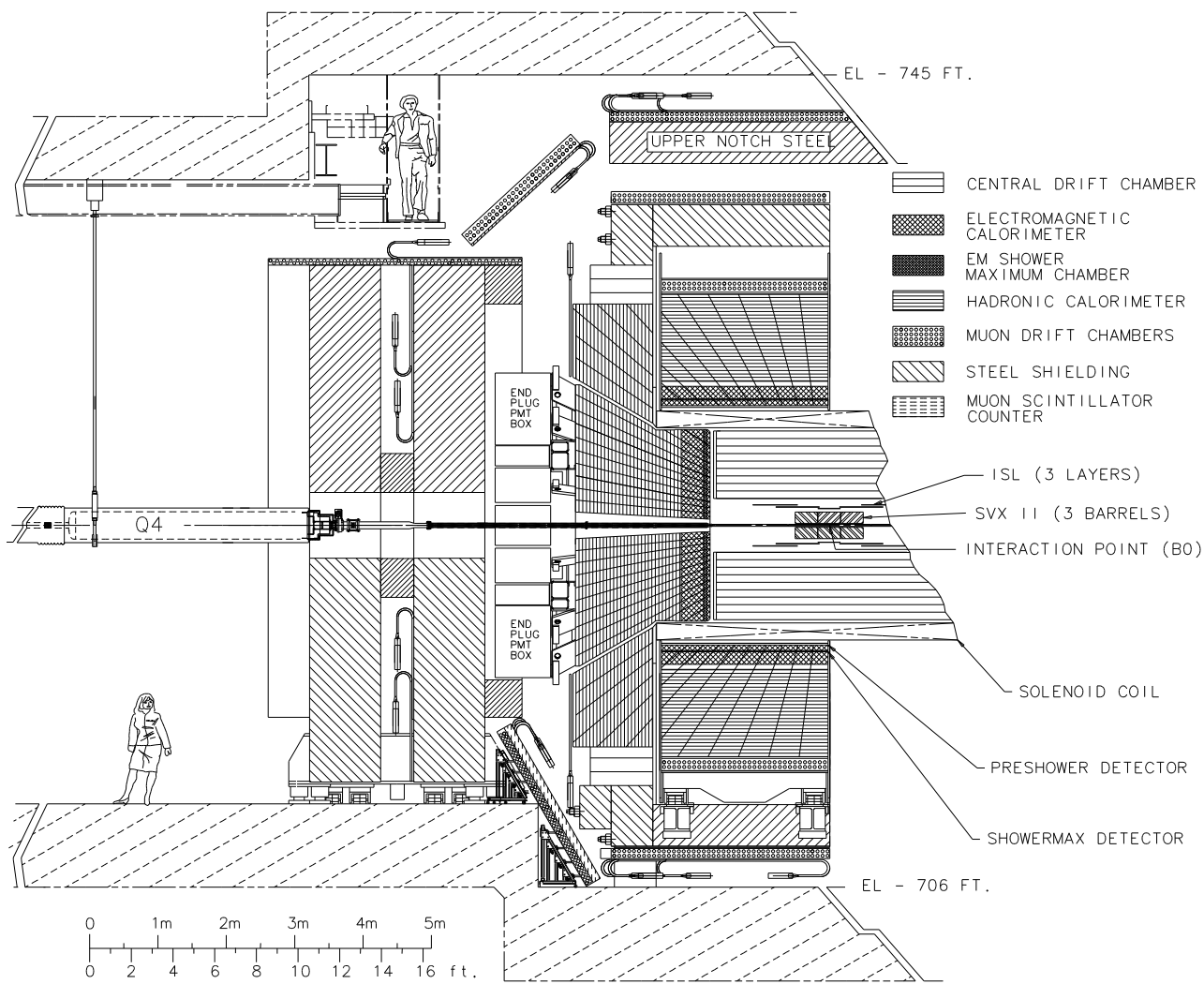}
  \caption{Elevation view of half of the CDF Run II detector.}
\label{fig:CDFIIdetector}
\end{figure}

\begin{figure}[p]
\centering
\includegraphics[width=\figurewidth]{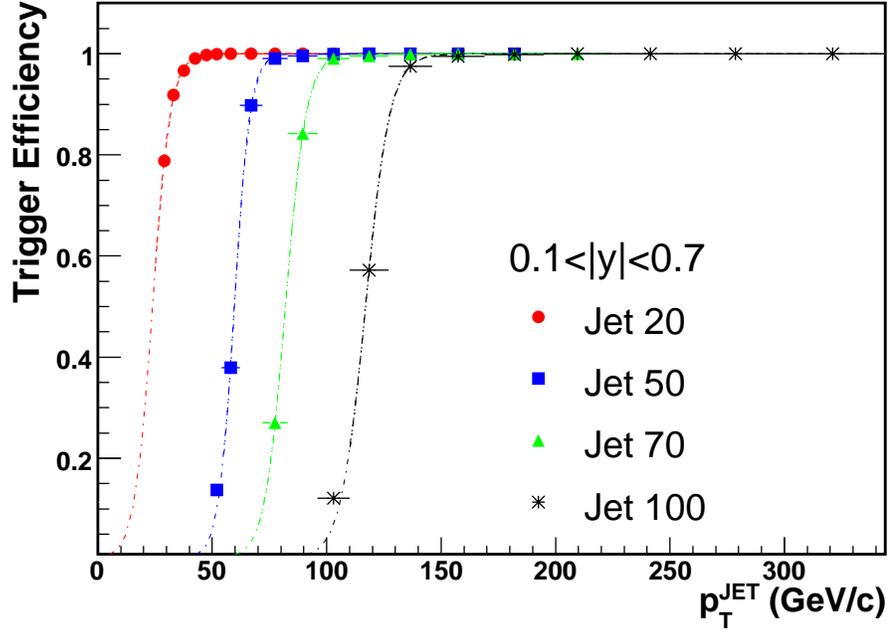}
 \caption{Jet trigger efficiencies as functions of jet \pt\
   for four trigger paths in the rapidity region $0.1<|y|<0.7$.
   The jet $p_T$ measured by the
   calorimeters is corrected
   as described in Sec.~\ref{sec:Corrections} in these
   distributions.}
\label{fig:trig_eff}
\end{figure}

\begin{figure}[p]
\centering
\includegraphics[width=\figurewidth]{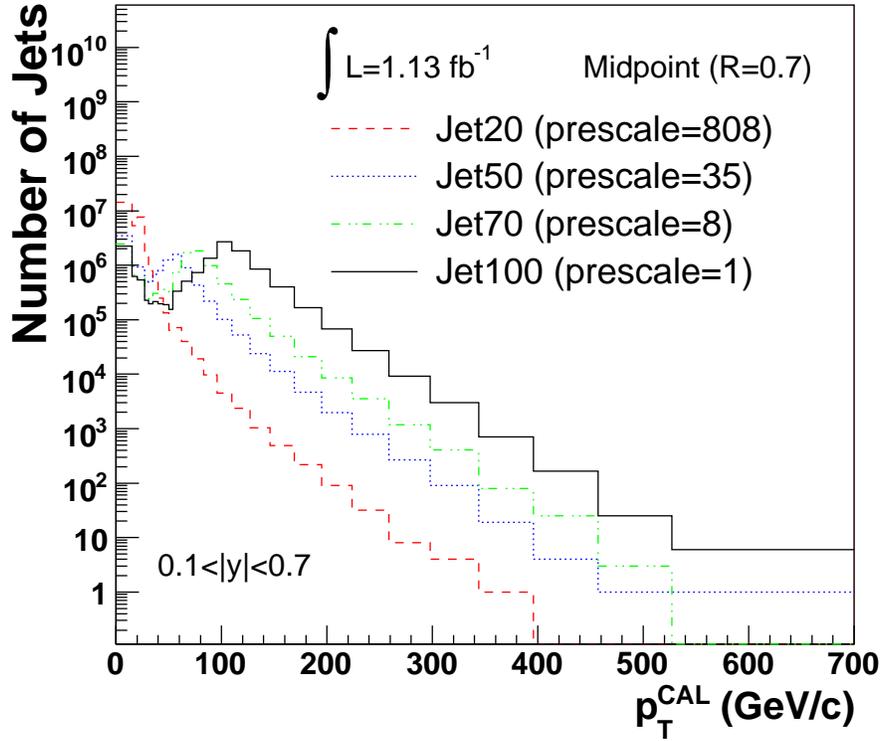}
\caption{Jet yield distributions as functions of jet \pt\ 
  for four trigger paths in the rapidity region $0.1<|y|<0.7$ with no
  correction for trigger prescales.}
\label{fig:jetpt}
\end{figure}

\begin{figure}[htb]
\centering
\includegraphics[width=\figurewidth,bb=150 510 470 740]{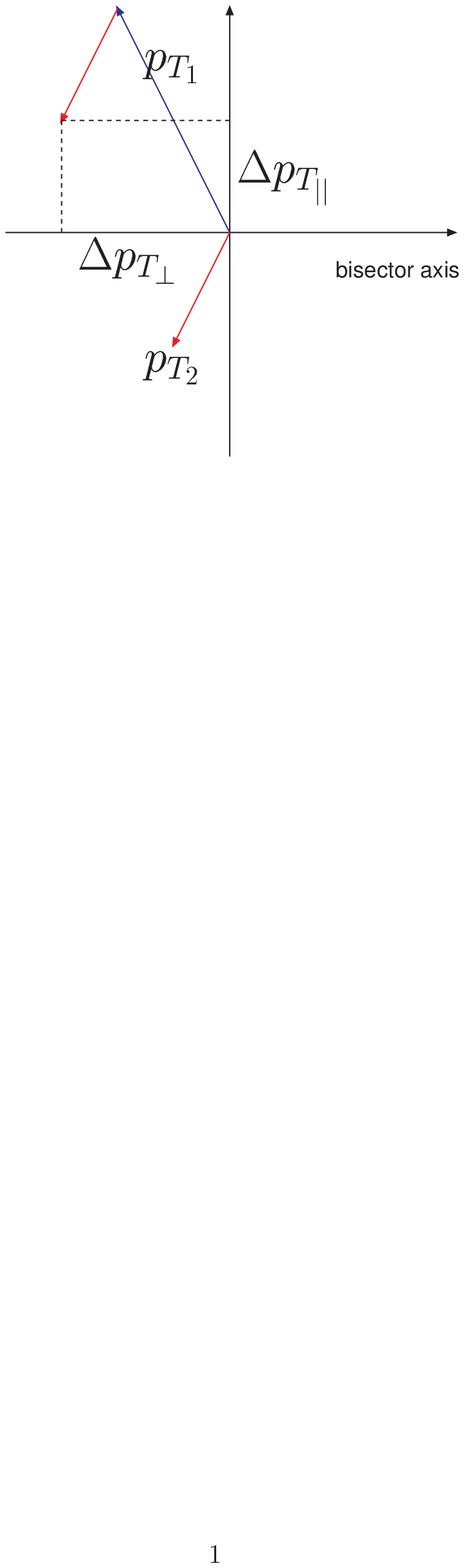}
\caption{The bisector variables described in the text are labeled in
  the diagram of the transverse plane shown above.  $\Delta
  p_{T_{\perp}}$ ($\Delta p_{T_{||}}$) is defined to be the
  component along (perpendicular to) the bisector axis of the sum of
  the jet $p_T$s.}
 \label{fig:bisector}
\end{figure}

\begin{figure}[p]
\centering
\includegraphics[width=\wfigurewidth]{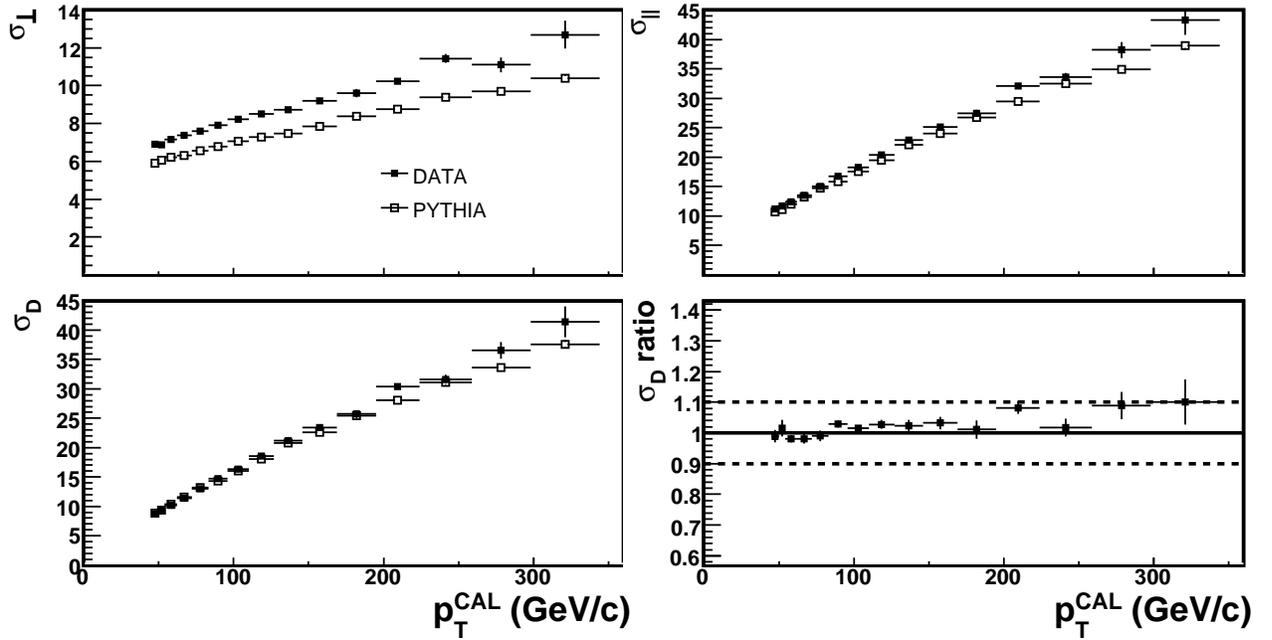}
 \caption{The $\sigma_{\perp}$, $\sigma_{||}$, and $\sigma_D$
   distributions as
   functions of jet $p_T$ in data and \py\ MC events in the
   region of $0.1<|y|<0.7$. The data/\py\ ratio of $\sigma_D$
   is also shown.} 
\label{fig:bisector_central}
\end{figure}

\begin{figure}[p]
\centering
\includegraphics[width=\wfigurewidth]{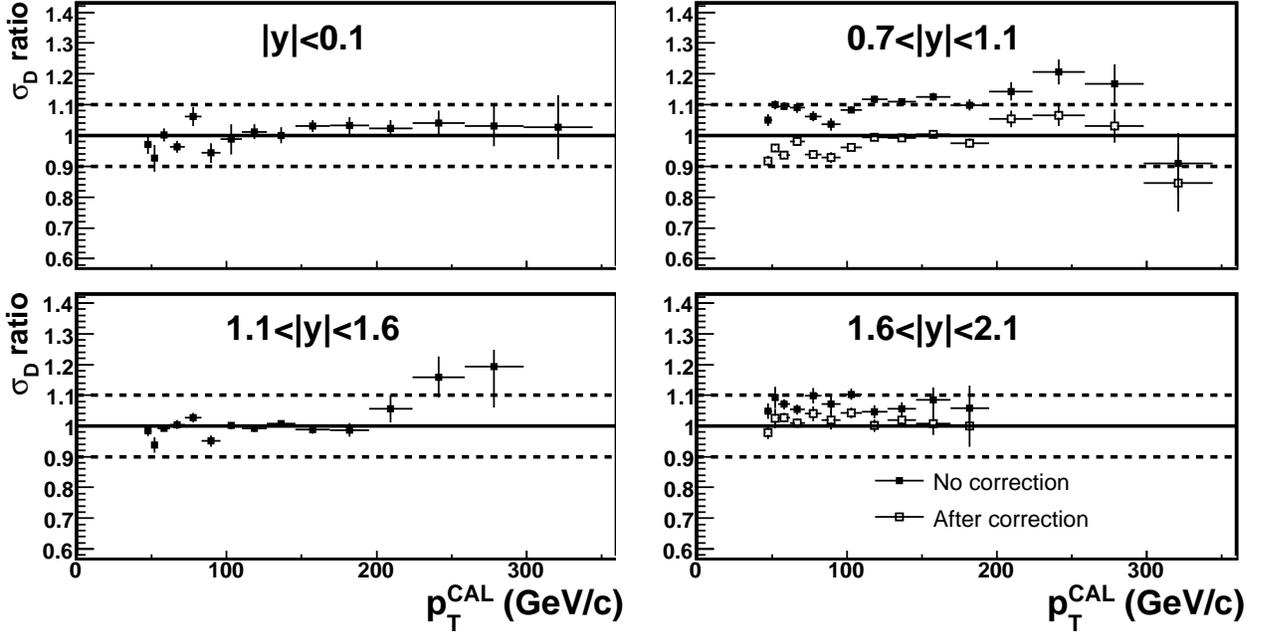}
 \caption{The data/MC ratio in $\sigma_{D}$ in four
 rapidity regions, $|y|<0.1$, $0.7<|y|<1.1$, $1.1<|y|<1.6$, and
 $1.6<|y|<2.1$ (solid squares).
 The same distribution in the region of $0.1<|y|<0.7$ is shown in
 Fig.~\ref{fig:bisector_central}.
 In the two regions where the simulation
 slightly underestimates the detector resolution, the MC is
 smeared by an additional amount to bring simulation and data into
 agreement (open squares).}   
\label{fig:bisector_all}
\end{figure}

\begin{figure}[p]
\centering
\includegraphics[width=\figurewidth]{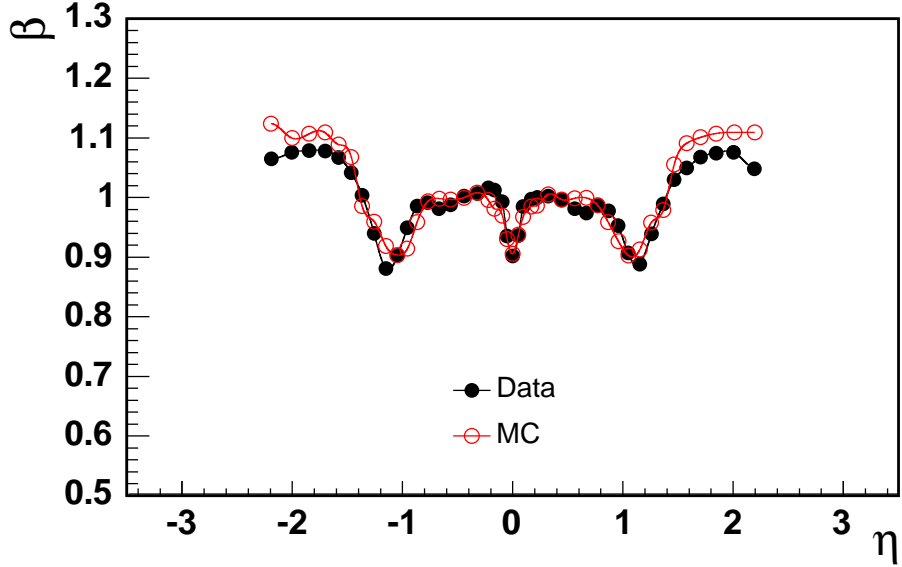}
 \caption{The dijet $p_T$ balance ($\beta$), as defined in
   Sec.~\ref{subsec:relcor}, 
   for data and {\sc pythia} MC events as a function of jet $\eta$.}   
\label{fig:relcor}
\end{figure}

\begin{figure}[p]
\centering
\includegraphics[width=\wfigurewidth]{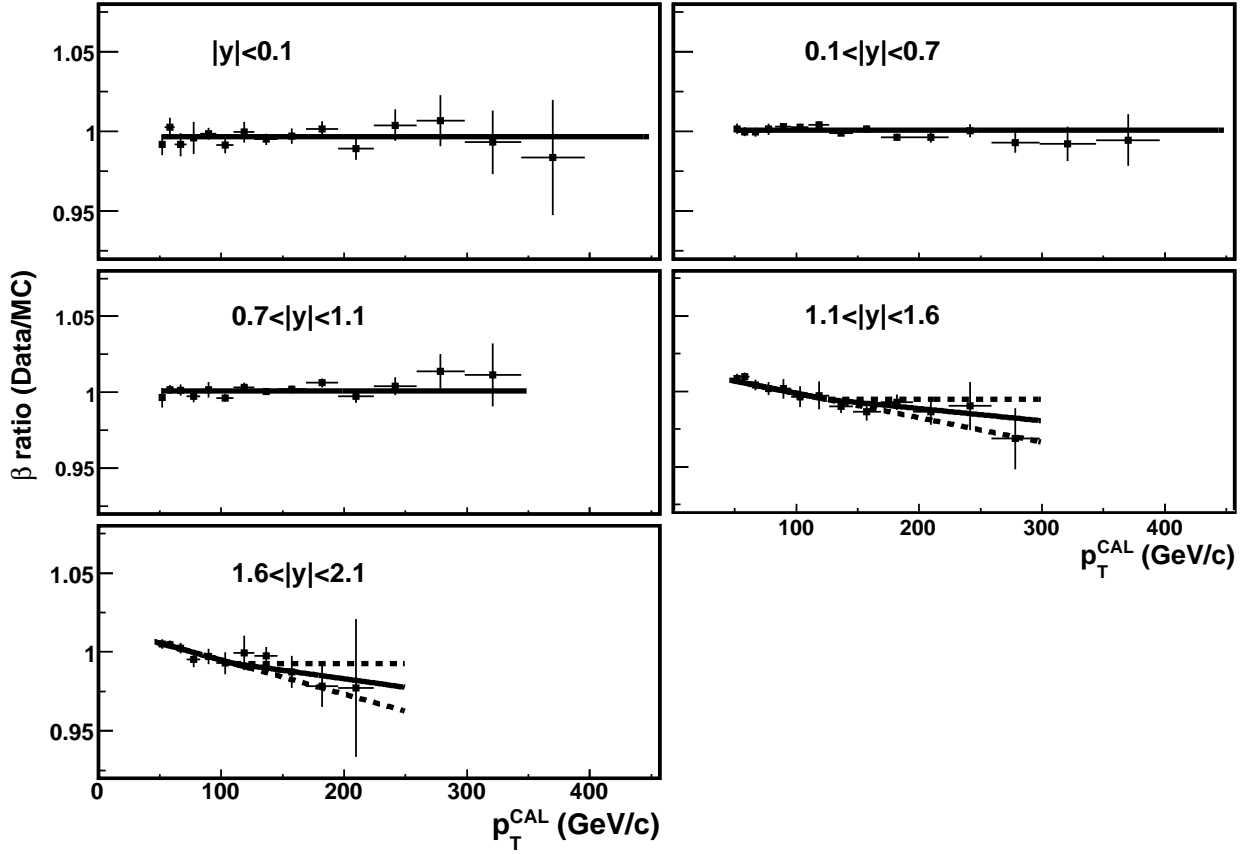}
 \caption{The data/MC ratio in dijet $p_T$ balance ($\beta$) in five
   rapidity regions after the $p_T$-independent relative correction have been
   applied. An additional correction is applied to the MC events to bring them
   into agreement with data in the regions of $|y|<0.1$,
   $1.1<|y|<1.6$, and $1.6<|y|<2.1$.
   In the most forward regions ($1.1<|y|<1.6$ and $1.6<|y|<2.1$),
   an additional systematic uncertainty is quoted due to the limited
   statistics at large $p_T$ which is shown by the the dashed lines.}
\label{fig:djbal}
\end{figure}

\begin{figure}[p]
\centering
\includegraphics[width=\wfigurewidth]{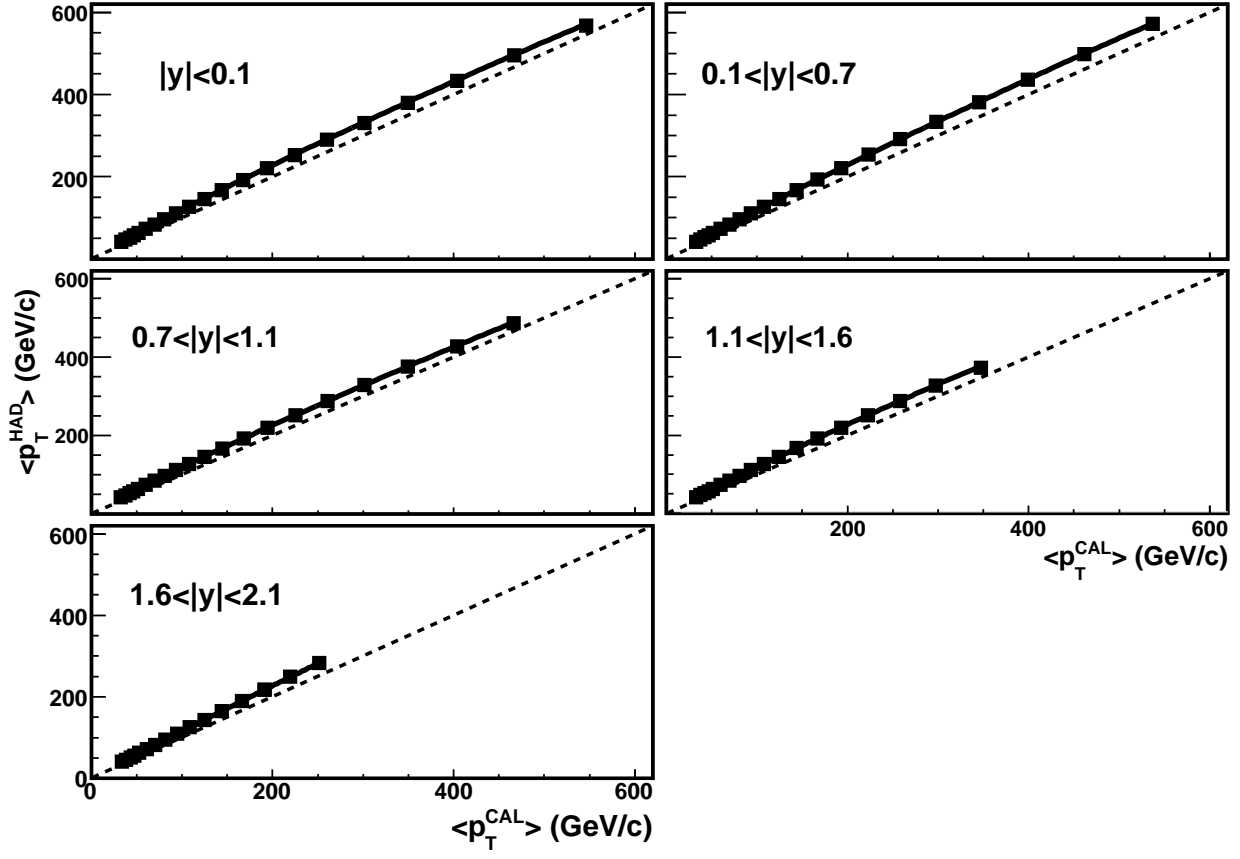}
 \caption{The average hadron level jet $p_T$ as a function of the
   average calorimeter jet \pt\ in five rapidity regions. The dashed
   lines correspond to 
   $\langle p_T^{\rm CAL} \rangle = \langle p_T^{\rm HAD} \rangle$.}
\label{fig:avgcor}
\end{figure}

\begin{figure}[p]
\centering
\includegraphics[width=\wfigurewidth]{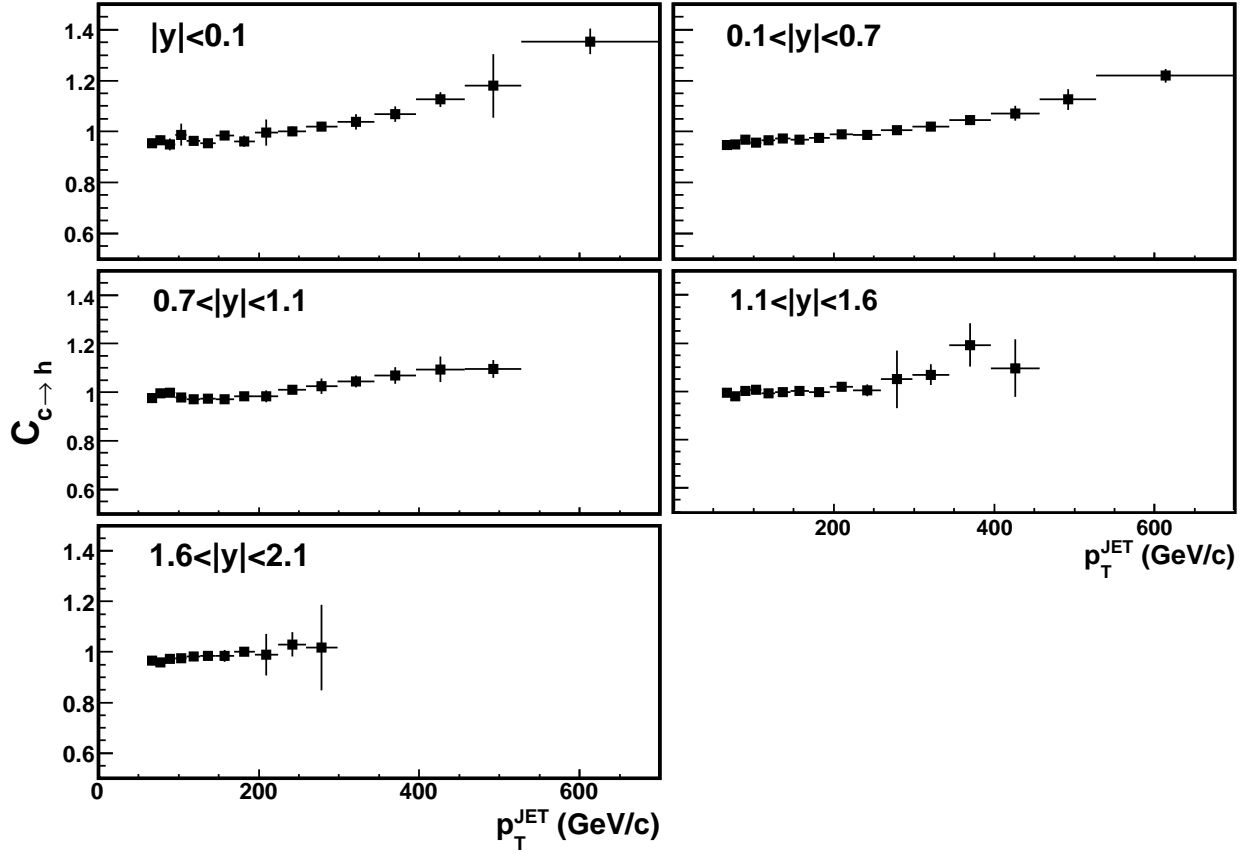}
 \caption{The unfolding correction factors as functions of jet $p_T$
   in five rapidity regions.}   
\label{fig:bincor}
\end{figure}

\begin{figure}[p]
\centering
\includegraphics[width=\wfigurewidth]{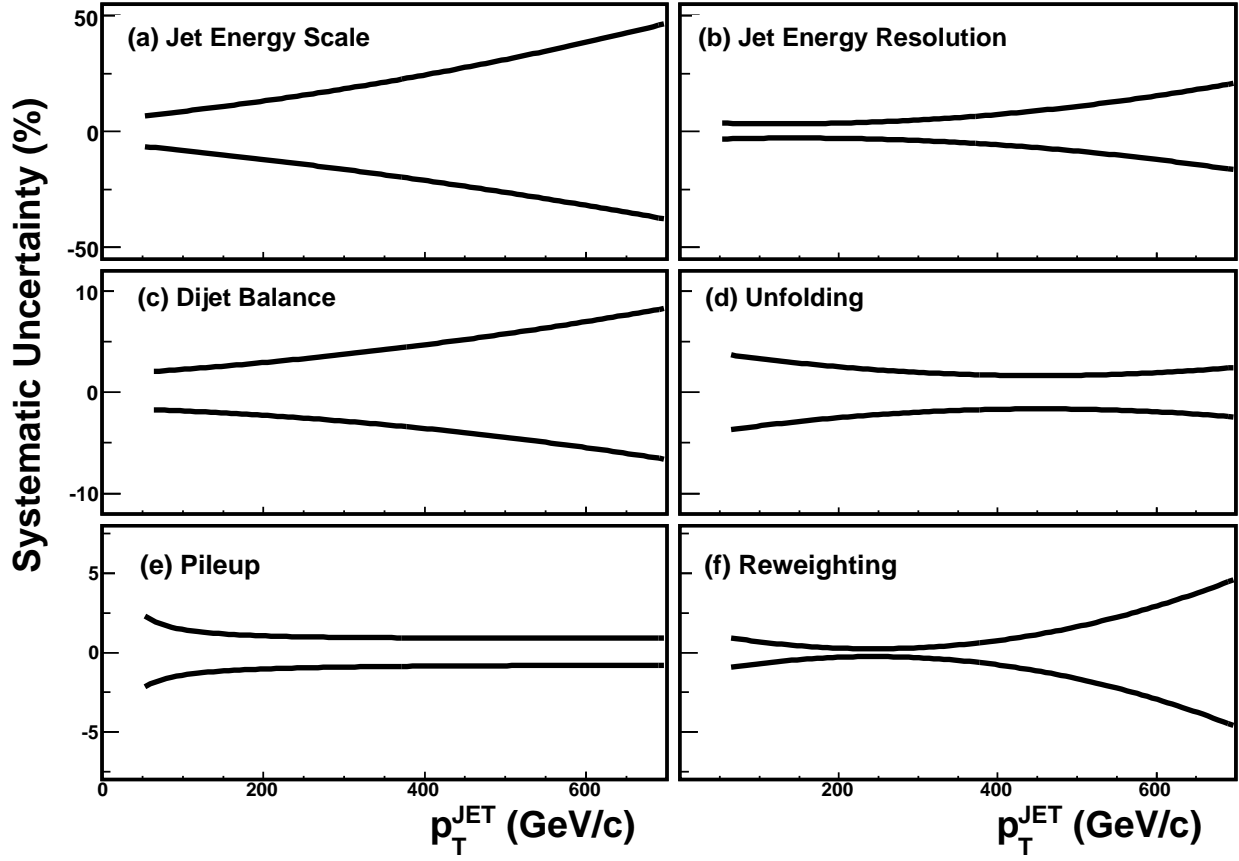}
\caption{The systematic uncertainty on the cross section
  in the rapidity region $0.1<|y|<0.7$ for each source considered in
  the measurement.}
\label{fig:sys_central}
\end{figure}

\begin{figure}[p]
\centering
\includegraphics[width=\sfigurewidth]{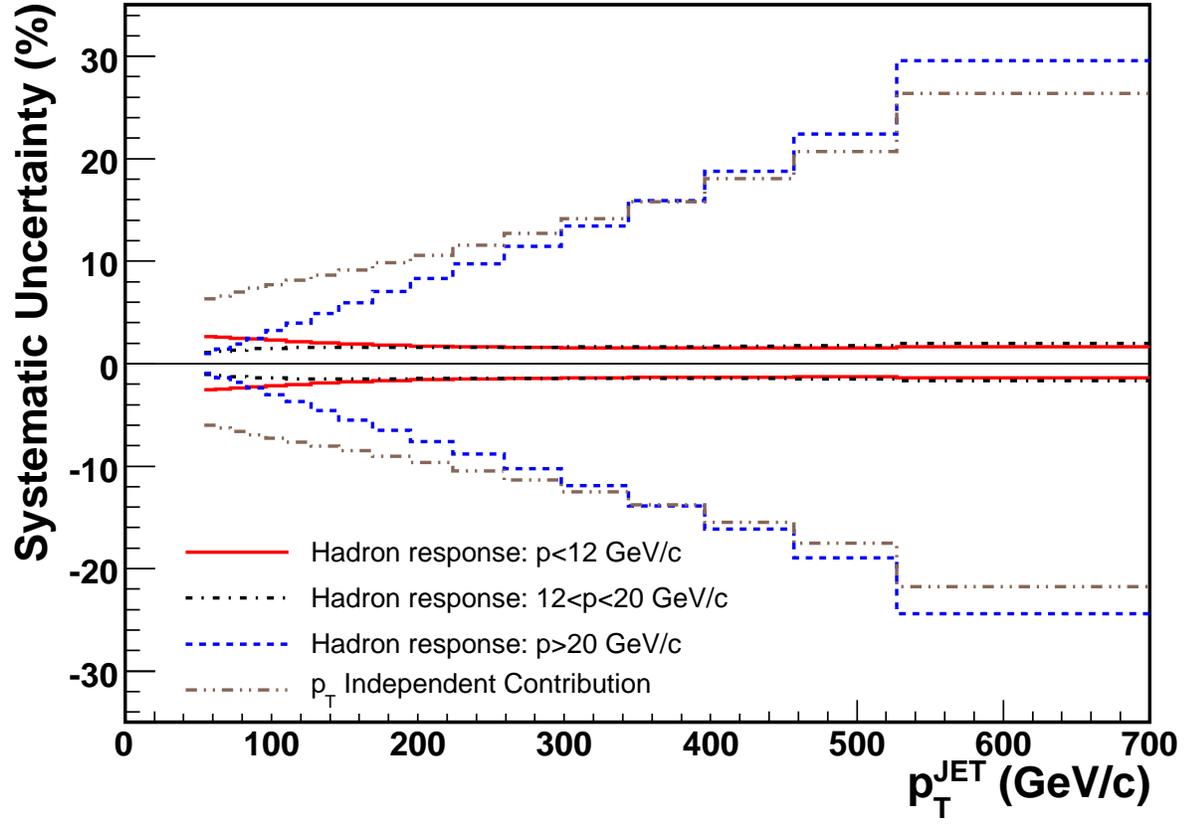}
\caption{The relative uncertainty on the jet cross section in the
  rapidity region $0.1<|y|<0.7$ due to different components of the jet
  energy scale systematic uncertainties.
  The decomposition includes contributions due to the description of
  the calorimeter response to hadrons for three different ranges of
  hadron momentum and a $p_T$-independent component as discussed in
  Sec.~\ref{sec:Systematics}.}
\label{fig:jes_breakdown}
\end{figure}

\begin{figure}[p]
\centering
\includegraphics[width=\wfigurewidth]{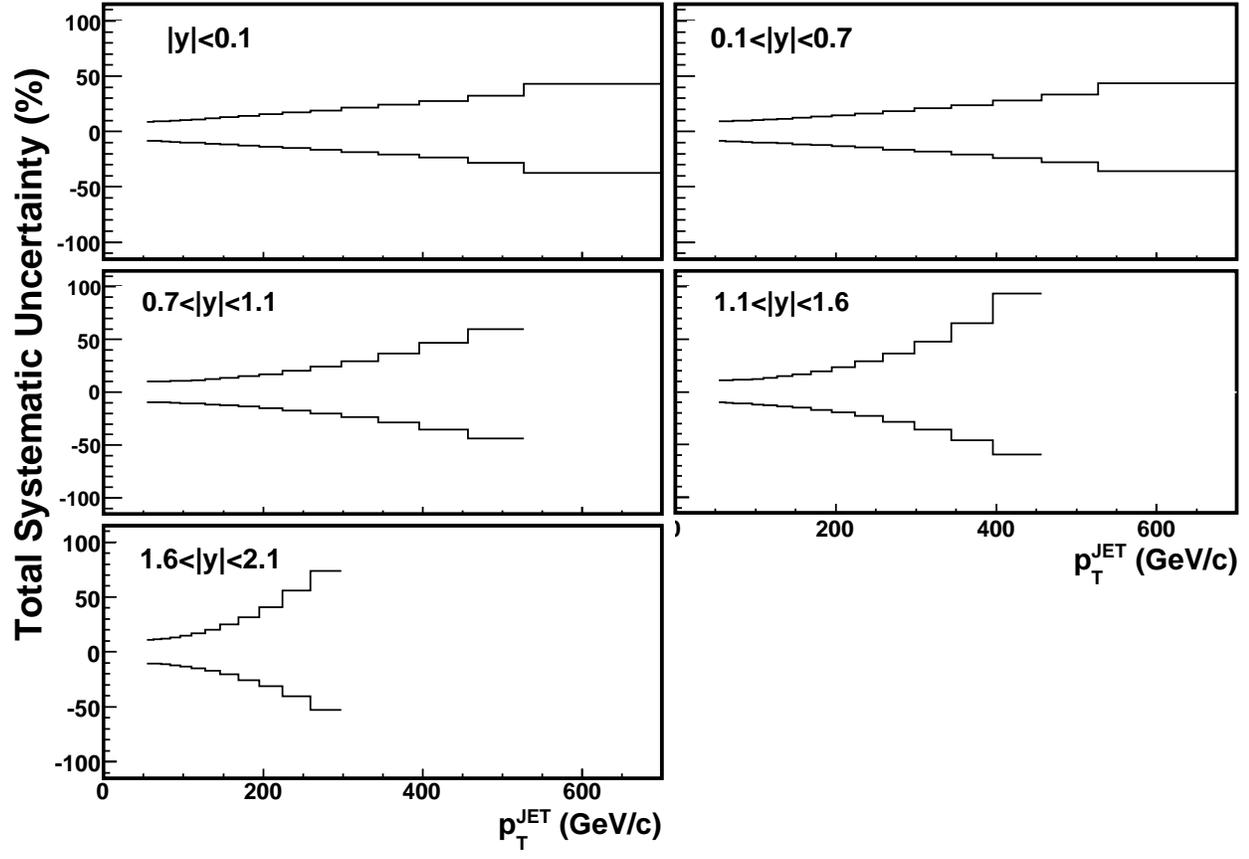}
\caption{The total systematic uncertainty on the cross
  section in five rapidity regions.
}
\label{fig:sys_tot}
\end{figure}

\begin{figure}[p]
\centering
\includegraphics[width=\wfigurewidth]{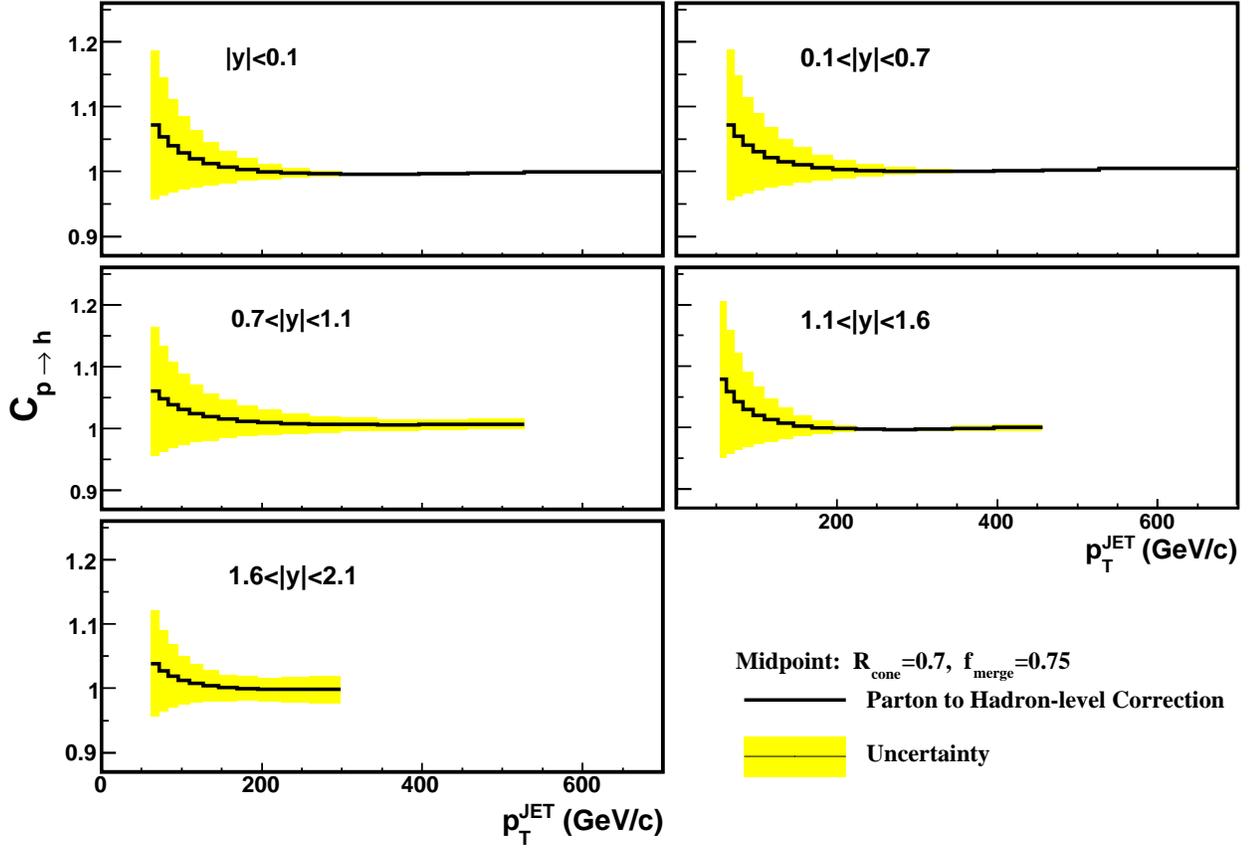}
\caption{The parton-to-hadron-level correction for
  five rapidity regions.  The correction is derived from \py\ (solid line)
  and the difference between the \hw\ and \py\ prediction for the
  correction is conservatively taken as
  the systematic uncertainty (shaded bands).
}
\label{fig:Par2Had_sys}
\end{figure}

\begin{figure}[p]
\centering
\includegraphics[width=\rfigurewidth]{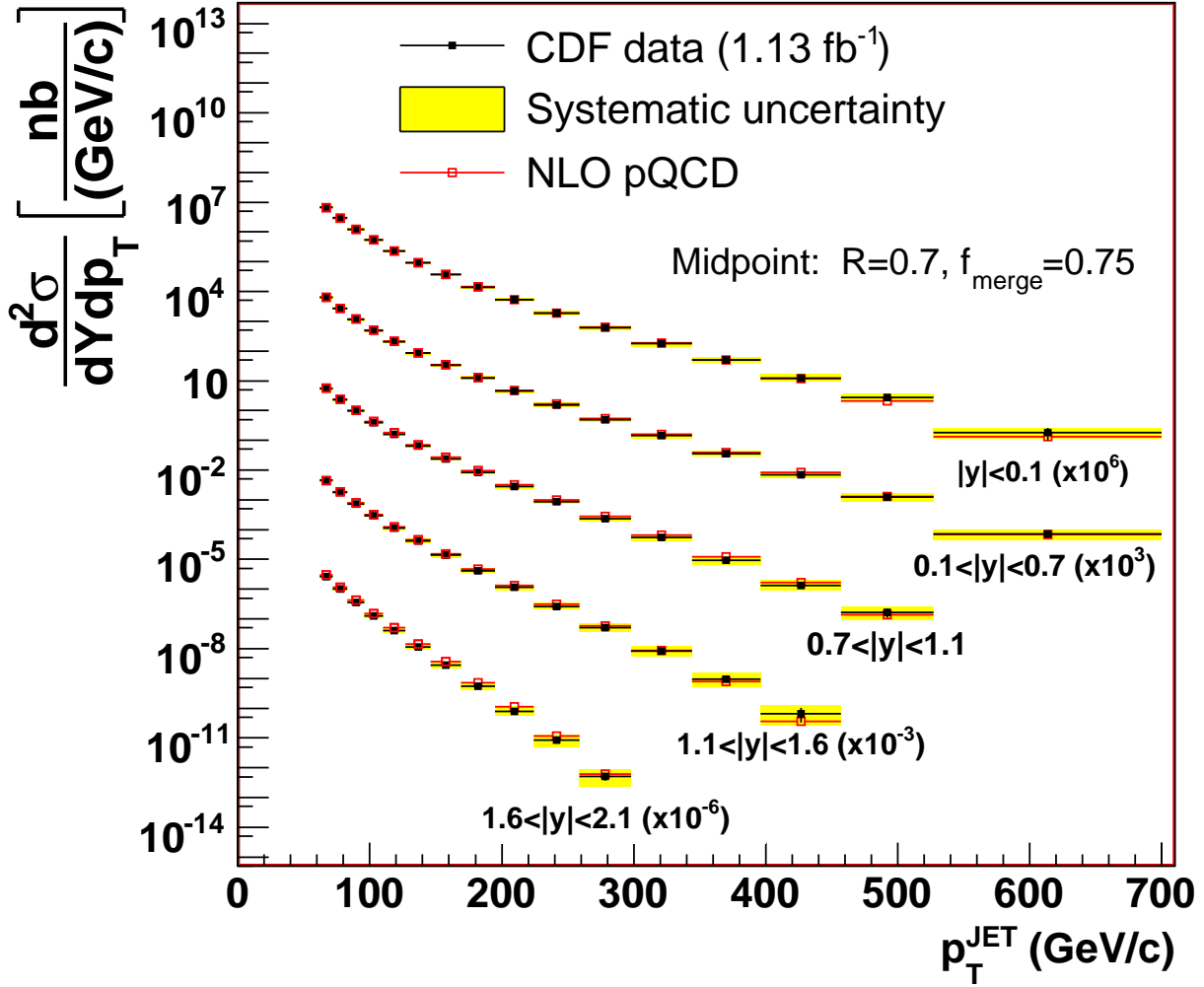}
\caption{Inclusive jet cross sections measured at the hadron level using
  the Midpoint algorithm in five rapidity regions compared to 
  NLO pQCD predictions based on the CTEQ6.1M PDF.
  The cross sections for the five rapidity regions are scaled
  by a factor of $10^3$ from each other for presentation purposes.}
\label{fig:result_dist_relcor_all}
\end{figure}

\begin{figure}[p]
\centering
\includegraphics[width=\rfigurewidth]{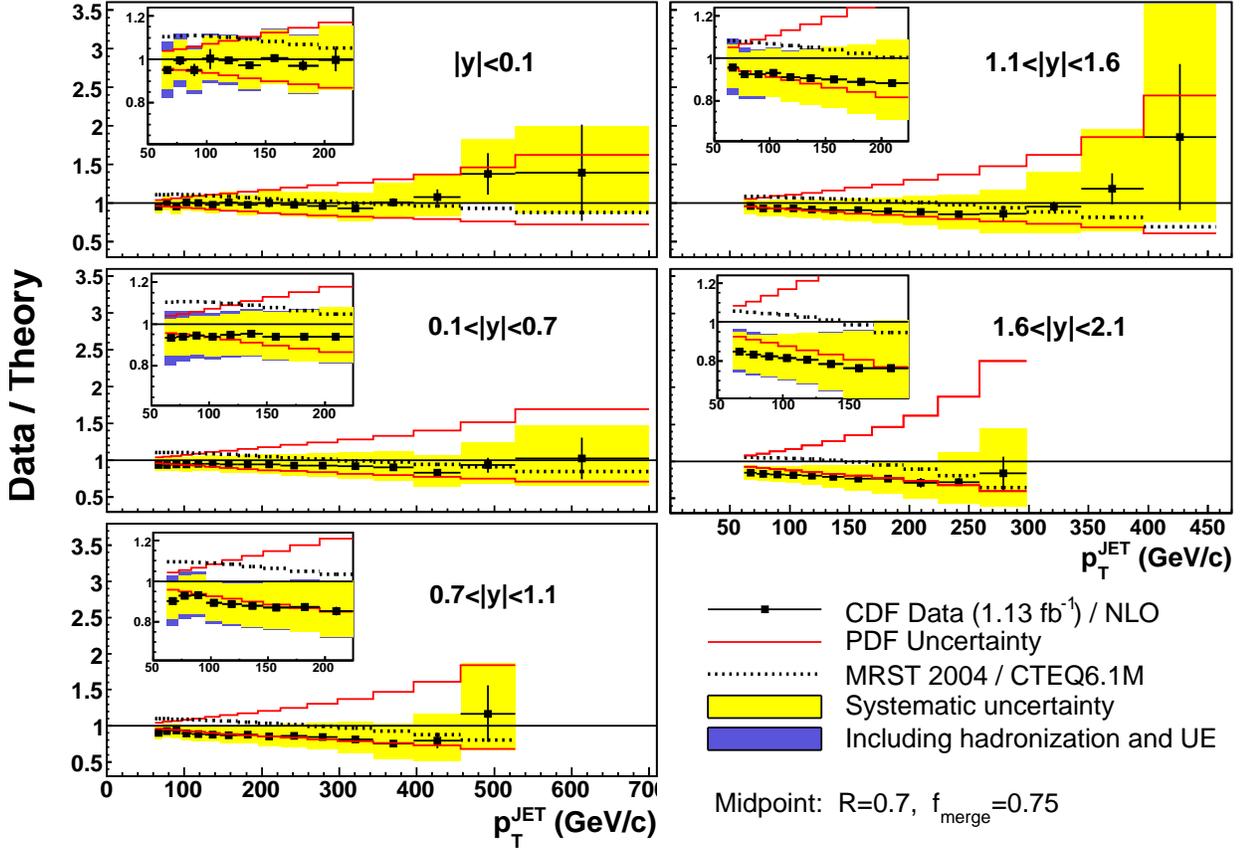}
\caption{The ratios of the measured inclusive jet cross sections at the
  hadron level with the Midpoint jet clustering algorithm to the NLO
  pQCD predictions (corrected to the hadron level) 
  in five rapidity regions.  
  Also shown are 
  the experimental systematic uncertainties on the measured cross section,
  the uncertainties in the hadronization and underlying event
  corrections added in quadrature with the experimental systematic uncertainties,
  and the PDF uncertainties on the theoretical predictions.
  The ratios of the theoretical predictions based on the MRST2004 and
  CTEQ6.1M are shown by the dotted lines.}
\label{fig:result_rat_relcor_all}
\end{figure}

\begin{figure}[p]
\centering
\includegraphics[width=\rfigurewidth]{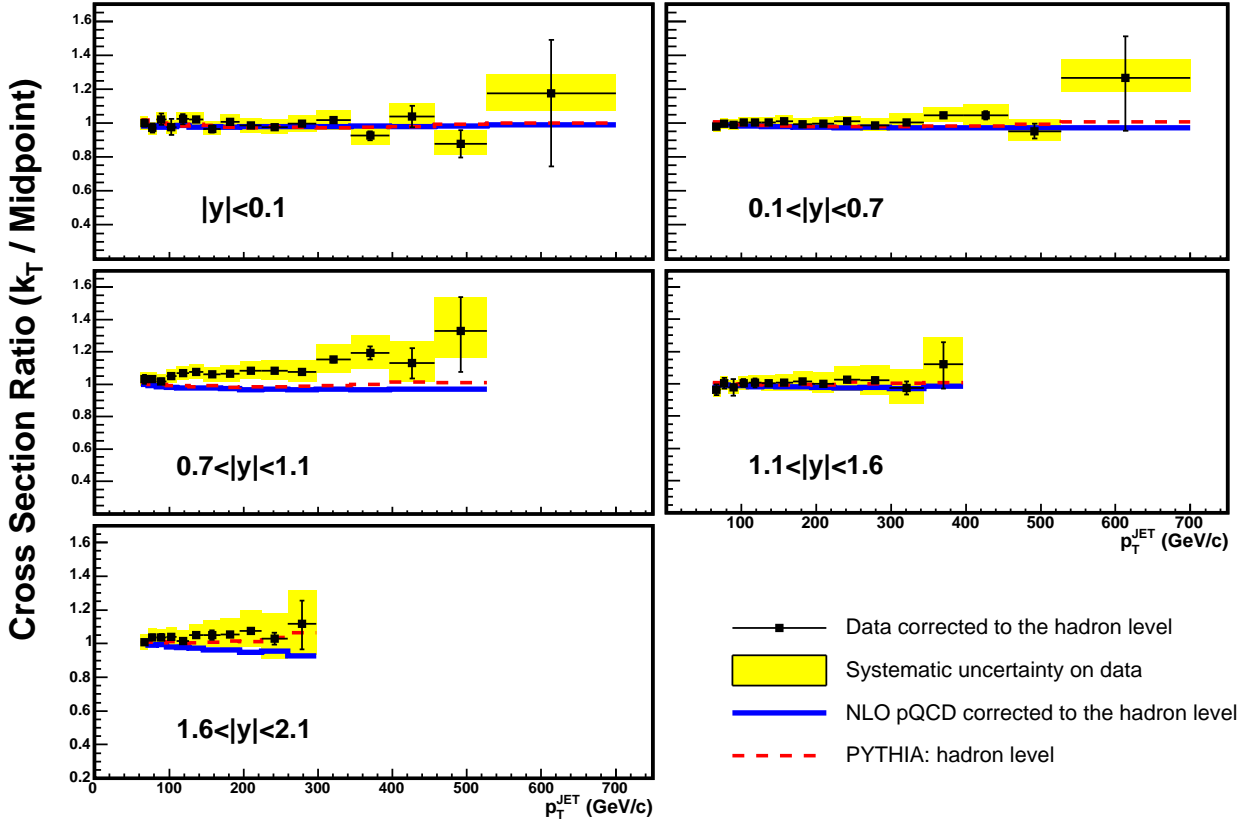}
\caption{The ratios of the inclusive jet cross sections measured
  using the $k_{T}$ algorithm with $D=0.7$~\cite{KT_PRD} to those measured
  using the Midpoint jet finding algorithm with $R_{\rm cone}=0.7$ in
  this paper (points).
  The systematic uncertainty on the ratio is given as the yellow band.
  The predictions from NLO pQCD (solid lines) and {\sc pythia} (dashed
  lines) for this ratio 
  are also shown.}
\label{fig:kt_mp_ratio_all}
\end{figure}

\clearpage

\begin{table*}[p] 
\caption{Relative contributions to the total jet energy scale uncertainty.}
\begin{ruledtabular} 
\begin{tabular}{ccccc} 
\multicolumn{1}{c}{$p_{T}$} & 
\multicolumn{1}{c}{~~$p_{T}$-independent~~} &
\multicolumn{3}{c}{response to hadrons} \\  
\multicolumn{1}{c}{~~(GeV/$c$)~~} &
\multicolumn{1}{c}{uncertainty} &
\multicolumn{1}{c}{~~$  p < 12$   GeV/$c$~~} &
\multicolumn{1}{c}{~~$12 <   p < 20$ GeV/$c$~~} &
\multicolumn{1}{c}{~~$  p > 20$   GeV/$c$~~} \\ \hline
 62 $-$ 72  & 90.0 & 35.1 & 16.7 & 19.7\\ 
 72 $-$ 83  & 89.8 & 32.1 & 17.3 & 24.8\\ 
 83 $-$ 96  & 89.1 & 29.1 & 17.7 & 30.1\\ 
 96 $-$ 110 & 87.7 & 26.1 & 16.9 & 36.6\\
110 $-$ 127 & 86.2 & 22.9 & 16.7 & 41.9\\
127 $-$ 146 & 84.1 & 19.9 & 15.4 & 47.9\\
146 $-$ 169 & 81.8 & 17.0 & 14.2 & 53.1\\
169 $-$ 195 & 79.7 & 14.8 & 12.9 & 57.1\\
195 $-$ 224 & 77.5 & 12.6 & 11.7 & 60.8\\
224 $-$ 259 & 75.6 & 10.9 & 10.5 & 63.7\\
259 $-$ 298 & 73.6 &  9.2 &  9.3 & 66.4\\ 
298 $-$ 344 & 71.9 &  7.9 &  8.3 & 68.5\\ 
344 $-$ 396 & 70.1 &  6.7 &  7.3 & 70.6\\ 
396 $-$ 457 & 69.0 &  5.9 &  6.5 & 71.9\\ 
457 $-$ 527 & 67.7 &  5.0 &  5.8 & 73.2\\ 
527 $-$ 700 & 66.4 &  4.2 &  5.0 & 74.5\\ 
\end{tabular}
\end{ruledtabular}  
\label{tab:decompose}   
\end{table*}

\begin{table*}[p]
\caption{Systematic uncertainties on the measured
  inclusive jet differential cross sections
  from different sources (as discussed in Sec.~\ref{sec:Systematics})
  as a function of \pt\ for jets in the region $|y|<0.1$.}
\begin{ruledtabular}
\begin{tabular}{cccccccc}
\multicolumn{8}{c}{\normalsize{Systematic uncertainties [$\%$]
    $(|y|<0.1)$}} \\ \hline
\multicolumn{2}{c}{} &
\multicolumn{2}{c}{dijet $p_T$ balance} &
\multicolumn{4}{c}{} \\
\multicolumn{1}{c}{\pt\ (GeV/$c$)} &
\multicolumn{1}{c}{jet energy scale} &
\multicolumn{1}{c}{nominal} &
\multicolumn{1}{c}{additional} &
\multicolumn{1}{c}{pileup} &
\multicolumn{1}{c}{unfolding} &
\multicolumn{1}{c}{$p_T$-spectra} &
\multicolumn{1}{c}{resolution} \\  \hline
 62.0 $-$  72.0 & ${}^{+ 6.7 }_{ -6.5 }$   & ${}^{+ 2.3 }_{ -1.7 }$ &---& ${}^{+ 1.9 }_{-1.9 }$ & $\pm~ 3.7 ~$ & $\pm~ 0.1 ~$  &${}^{+ 3.9 }_{ -3.6 }$   \\
 72.0 $-$  83.0 & ${}^{+ 7.3 }_{ -7.0 }$   & ${}^{+ 1.9 }_{ -2.0 }$ &---& ${}^{+ 1.8 }_{-1.7 }$ & $\pm~ 3.6 ~$ & $\pm~ 0.2 ~$  &${}^{+ 3.9 }_{ -3.6 }$   \\
 83.0 $-$  96.0 & ${}^{+ 7.9 }_{ -7.5 }$   & ${}^{+ 2.3 }_{ -1.7 }$ &---& ${}^{+ 1.7 }_{-1.6 }$ & $\pm~ 3.5 ~$ & $\pm~ 1.0 ~$  &${}^{+ 3.9 }_{ -3.6 }$   \\
 96.0 $-$ 110.0 & ${}^{+ 8.6 }_{ -8.1 }$   & ${}^{+ 2.4 }_{ -2.3 }$ &---& ${}^{+ 1.6 }_{-1.4 }$ & $\pm~ 3.3 ~$ & $\pm~ 0.6 ~$  &${}^{+ 4.0 }_{ -3.6 }$   \\
110.0 $-$ 127.0 & ${}^{+ 9.5 }_{ -8.7 }$   & ${}^{+ 2.0 }_{ -1.8 }$ &---& ${}^{+ 1.5 }_{-1.3 }$ & $\pm~ 3.2 ~$ & $\pm~ 0.7 ~$  &${}^{+ 4.0 }_{ -3.6 }$   \\
127.0 $-$ 146.0 & ${}^{+ 10.4 }_{ -9.5 }$  & ${}^{+ 2.6 }_{ -2.4 }$ &---& ${}^{+ 1.3 }_{-1.2 }$ & $\pm~ 3.1 ~$ & $\pm~ 1.5 ~$  &${}^{+ 4.1 }_{ -3.7 }$   \\
146.0 $-$ 169.0 & ${}^{+ 11.5 }_{ -10.4 }$ & ${}^{+ 2.6 }_{ -1.4 }$ &---& ${}^{+ 1.2 }_{-1.1 }$ & $\pm~ 2.9 ~$ & $\pm~ 1.1 ~$  &${}^{+ 4.2 }_{ -3.8 }$   \\
169.0 $-$ 195.0 & ${}^{+ 12.8 }_{ -11.5 }$ & ${}^{+ 3.1 }_{ -2.7 }$ &---& ${}^{+ 1.1 }_{-1.1 }$ & $\pm~ 2.7 ~$ & $\pm~ 1.6 ~$  &${}^{+ 4.5 }_{ -3.9 }$   \\
195.0 $-$ 224.0 & ${}^{+ 14.2 }_{ -12.6 }$ & ${}^{+ 3.2 }_{ -2.5 }$ &---& ${}^{+ 0.9 }_{-1.0 }$ & $\pm~ 2.5 ~$ & $\pm~ 1.2 ~$  &${}^{+ 4.8 }_{ -4.1 }$   \\
224.0 $-$ 259.0 & ${}^{+ 15.9 }_{ -14.0 }$ & ${}^{+ 2.8 }_{ -2.3 }$ &---& ${}^{+ 0.8 }_{-1.0 }$ & $\pm~ 2.3 ~$ & $\pm~ 1.6 ~$  &${}^{+ 5.2 }_{ -4.4 }$   \\
259.0 $-$ 298.0 & ${}^{+ 17.8 }_{ -15.5 }$ & ${}^{+ 3.4 }_{ -2.4 }$ &---& ${}^{+ 0.7 }_{-0.9 }$ & $\pm~ 2.1 ~$ & $\pm~ 0.8 ~$  &${}^{+ 5.8 }_{ -4.9 }$   \\
298.0 $-$ 344.0 & ${}^{+ 19.9 }_{ -17.3 }$ & ${}^{+ 4.2 }_{ -3.4 }$ &---& ${}^{+ 0.5 }_{-0.9 }$ & $\pm~ 1.9 ~$ & $\pm~ 0.2 ~$  &${}^{+ 6.6 }_{ -5.5 }$   \\
344.0 $-$ 396.0 & ${}^{+ 22.4 }_{ -19.3 }$ & ${}^{+ 4.0 }_{ -2.9 }$ &---& ${}^{+ 0.4 }_{-0.9 }$ & $\pm~ 1.7 ~$ & $\pm~ 0.4 ~$  &${}^{+ 7.8 }_{ -6.4 }$   \\
396.0 $-$ 457.0 & ${}^{+ 25.1 }_{ -21.6 }$ & ${}^{+ 5.1 }_{ -4.3 }$ &---& ${}^{+ 0.3 }_{-0.9 }$ & $\pm~ 1.5 ~$ & $\pm~ 2.2 ~$  &${}^{+ 9.5 }_{ -7.7 }$   \\
457.0 $-$ 527.0 & ${}^{+ 28.3 }_{ -24.2 }$ & ${}^{+ 5.7 }_{ -4.7 }$ &---& ${}^{+ 0.1 }_{-0.8 }$ & $\pm~ 1.4 ~$ & $\pm~ 9.3 ~$  &${}^{+ 11.7 }_{ -9.5 }$  \\
527.0 $-$ 700.0 & ${}^{+ 34.0 }_{ -29.0 }$ & ${}^{+ 6.0 }_{ -4.8 }$ &---& ${}^{+ 0.1 }_{-0.8 }$ & $\pm~ 1.4 ~$ & $\pm~ 19.1 ~$ &${}^{+ 16.9 }_{ -13.7 }$ \\
\end{tabular}
\end{ruledtabular}
\label{tab:sys1}
\end{table*}

\begin{table*}[p]
\caption{Systematic uncertainties on the measured
  inclusive jet differential cross section 
  from different sources (as discussed in Sec.~\ref{sec:Systematics})
  as a function of \pt\ for jets in the region $0.1<|y|<0.7$.}
\begin{ruledtabular}
\begin{tabular}{cccccccc}
\multicolumn{8}{c}{\normalsize{Systematic uncertainties [$\%$]
    $(0.1<|y|<0.7)$}} \\ \hline
\multicolumn{2}{c}{} &
\multicolumn{2}{c}{dijet $p_T$ balance} &
\multicolumn{4}{c}{} \\
\multicolumn{1}{c}{\pt\ (GeV/$c$)} &
\multicolumn{1}{c}{jet energy scale} &
\multicolumn{1}{c}{nominal} &
\multicolumn{1}{c}{additional} &
\multicolumn{1}{c}{pileup} &
\multicolumn{1}{c}{unfolding} &
\multicolumn{1}{c}{$p_T$-spectra} &
\multicolumn{1}{c}{resolution} \\  \hline
 62.0 $-$  72.0 & ${}^{+ 7.4 }_{ -7.0 }$   & ${}^{+ 2.2 }_{ -1.8 }$ &---& ${}^{+ 1.9 }_{-1.8 }$ & $\pm~ 3.7 ~$ & $\pm~ 0.5 ~$ &${}^{+ 3.5 }_{ -3.1 }$   \\
 72.0 $-$  83.0 & ${}^{+ 7.8 }_{ -7.4 }$   & ${}^{+ 2.3 }_{ -1.9 }$ &---& ${}^{+ 1.7 }_{-1.7 }$ & $\pm~ 3.5 ~$ & $\pm~ 0.5 ~$ &${}^{+ 3.4 }_{ -3.0 }$   \\
 83.0 $-$  96.0 & ${}^{+ 8.3 }_{ -7.8 }$   & ${}^{+ 2.2 }_{ -1.9 }$ &---& ${}^{+ 1.6 }_{-1.5 }$ & $\pm~ 3.4 ~$ & $\pm~ 0.5 ~$ &${}^{+ 3.4 }_{ -3.0 }$   \\
 96.0 $-$ 110.0 & ${}^{+ 8.8 }_{ -8.3 }$   & ${}^{+ 2.4 }_{ -2.3 }$ &---& ${}^{+ 1.4 }_{-1.4 }$ & $\pm~ 3.3 ~$ & $\pm~ 0.7 ~$ &${}^{+ 3.3 }_{ -2.9 }$   \\
110.0 $-$ 127.0 & ${}^{+ 9.5 }_{ -8.9 }$   & ${}^{+ 2.1 }_{ -1.8 }$ &---& ${}^{+ 1.3 }_{-1.3 }$ & $\pm~ 3.2 ~$ & $\pm~ 0.5 ~$ &${}^{+ 3.3 }_{ -2.9 }$   \\
127.0 $-$ 146.0 & ${}^{+ 10.3 }_{ -9.6 }$  & ${}^{+ 2.7 }_{ -1.9 }$ &---& ${}^{+ 1.2 }_{-1.2 }$ & $\pm~ 3.0 ~$ & $\pm~ 0.6 ~$ &${}^{+ 3.3 }_{ -2.8 }$   \\
146.0 $-$ 169.0 & ${}^{+ 11.2 }_{ -10.4 }$ & ${}^{+ 2.8 }_{ -2.4 }$ &---& ${}^{+ 1.2 }_{-1.1 }$ & $\pm~ 2.8 ~$ & $\pm~ 0.6 ~$ &${}^{+ 3.4 }_{ -2.8 }$   \\
169.0 $-$ 195.0 & ${}^{+ 12.3 }_{ -11.4 }$ & ${}^{+ 3.1 }_{ -2.3 }$ &---& ${}^{+ 1.1 }_{-1.1 }$ & $\pm~ 2.6 ~$ & $\pm~ 0.7 ~$ &${}^{+ 3.5 }_{ -2.9 }$   \\
195.0 $-$ 224.0 & ${}^{+ 13.7 }_{ -12.5 }$ & ${}^{+ 3.5 }_{ -2.7 }$ &---& ${}^{+ 1.0 }_{-1.0 }$ & $\pm~ 2.5 ~$ & $\pm~ 0.4 ~$ &${}^{+ 3.7 }_{ -3.0 }$   \\
224.0 $-$ 259.0 & ${}^{+ 15.3 }_{ -13.8 }$ & ${}^{+ 2.7 }_{ -2.1 }$ &---& ${}^{+ 1.0 }_{-1.0 }$ & $\pm~ 2.3 ~$ & $\pm~ 0.4 ~$ &${}^{+ 4.1 }_{ -3.2 }$   \\
259.0 $-$ 298.0 & ${}^{+ 17.3 }_{ -15.4 }$ & ${}^{+ 3.7 }_{ -2.8 }$ &---& ${}^{+ 1.0 }_{-0.9 }$ & $\pm~ 2.1 ~$ & $\pm~ 0.4 ~$ &${}^{+ 4.6 }_{ -3.6 }$   \\
298.0 $-$ 344.0 & ${}^{+ 19.6 }_{ -17.4 }$ & ${}^{+ 4.4 }_{ -3.2 }$ &---& ${}^{+ 1.0 }_{-0.9 }$ & $\pm~ 1.9 ~$ & $\pm~ 0.2 ~$ &${}^{+ 5.4 }_{ -4.2 }$   \\
344.0 $-$ 396.0 & ${}^{+ 22.6 }_{ -19.7 }$ & ${}^{+ 4.6 }_{ -3.5 }$ &---& ${}^{+ 0.9 }_{-0.9 }$ & $\pm~ 1.7 ~$ & $\pm~ 0.2 ~$ &${}^{+ 6.6 }_{ -5.1 }$   \\
396.0 $-$ 457.0 & ${}^{+ 26.1 }_{ -22.5 }$ & ${}^{+ 5.8 }_{ -4.5 }$ &---& ${}^{+ 0.9 }_{-0.8 }$ & $\pm~ 1.7 ~$ & $\pm~ 0.8 ~$ &${}^{+ 8.2 }_{ -6.4 }$   \\
457.0 $-$ 527.0 & ${}^{+ 30.6 }_{ -25.9 }$ & ${}^{+ 7.0 }_{ -5.4 }$ &---& ${}^{+ 0.9 }_{-0.8 }$ & $\pm~ 1.7 ~$ & $\pm~ 1.0 ~$ &${}^{+ 10.6 }_{ -8.2 }$  \\
527.0 $-$ 700.0 & ${}^{+ 39.7 }_{ -32.7 }$ & ${}^{+ 7.1 }_{ -5.5 }$ &---& ${}^{+ 0.9 }_{-0.8 }$ & $\pm~ 2.0 ~$ & $\pm~ 3.5 ~$ &${}^{+ 16.2 }_{ -12.8 }$ \\
\end{tabular}
\label{tab:sys2}
\end{ruledtabular}
\end{table*}

\begin{table*}[p]
\caption{Systematic uncertainties on the measured
  inclusive jet differential cross section 
  from different sources (as discussed in Sec.~\ref{sec:Systematics})
  as a function of \pt\ for jets in the region $0.7<|y|<1.1$.}
\begin{ruledtabular}
\begin{tabular}{cccccccc}
\multicolumn{8}{c}{\normalsize{Systematic uncertainties [$\%$]
    $(0.7<|y|<1.1)$}} \\ \hline
\multicolumn{2}{c}{} &
\multicolumn{2}{c}{dijet $p_T$ balance} &
\multicolumn{4}{c}{} \\
\multicolumn{1}{c}{\pt\ (GeV/$c$)} &
\multicolumn{1}{c}{jet energy scale} &
\multicolumn{1}{c}{nominal} &
\multicolumn{1}{c}{additional} &
\multicolumn{1}{c}{pileup} &
\multicolumn{1}{c}{unfolding} &
\multicolumn{1}{c}{$p_T$-spectra} &
\multicolumn{1}{c}{resolution} \\  \hline
 62.0 $-$  72.0 & ${}^{+ 8.0 }_{ -7.4 }$   & ${}^{+ 2.1 }_{ -1.9 }$  &---& ${}^{+ 2.0 }_{-1.9 }$ & $\pm~ 4.2 ~$ & $\pm~ 0.7 ~$ &${}^{+ 4.0 }_{ -3.3 }$   \\
 72.0 $-$  83.0 & ${}^{+ 8.3 }_{ -7.8 }$   & ${}^{+ 2.5 }_{ -2.0 }$  &---& ${}^{+ 1.8 }_{-1.7 }$ & $\pm~ 4.1 ~$ & $\pm~ 0.4 ~$ &${}^{+ 3.9 }_{ -3.3 }$   \\
 83.0 $-$  96.0 & ${}^{+ 8.7 }_{ -8.2 }$   & ${}^{+ 2.7 }_{ -2.0 }$  &---& ${}^{+ 1.7 }_{-1.6 }$ & $\pm~ 3.9 ~$ & $\pm~ 0.7 ~$ &${}^{+ 3.8 }_{ -3.2 }$   \\
 96.0 $-$ 110.0 & ${}^{+ 9.2 }_{ -8.7 }$   & ${}^{+ 2.0 }_{ -2.2 }$  &---& ${}^{+ 1.5 }_{-1.5 }$ & $\pm~ 3.8 ~$ & $\pm~ 0.6 ~$ &${}^{+ 3.8 }_{ -3.2 }$   \\
110.0 $-$ 127.0 & ${}^{+ 9.9 }_{ -9.3 }$   & ${}^{+ 2.2 }_{ -1.7 }$  &---& ${}^{+ 1.4 }_{-1.4 }$ & $\pm~ 3.6 ~$ & $\pm~ 0.7 ~$ &${}^{+ 3.8 }_{ -3.2 }$   \\
127.0 $-$ 146.0 & ${}^{+ 10.8 }_{ -10.1 }$ & ${}^{+ 2.9 }_{ -2.1 }$  &---& ${}^{+ 1.3 }_{-1.3 }$ & $\pm~ 3.4 ~$ & $\pm~ 0.7 ~$ &${}^{+ 3.9 }_{ -3.2 }$   \\
146.0 $-$ 169.0 & ${}^{+ 12.0 }_{ -11.1 }$ & ${}^{+ 3.1 }_{ -2.3 }$  &---& ${}^{+ 1.2 }_{-1.2 }$ & $\pm~ 3.2 ~$ & $\pm~ 0.7 ~$ &${}^{+ 4.1 }_{ -3.4 }$   \\
169.0 $-$ 195.0 & ${}^{+ 13.7 }_{ -12.4 }$ & ${}^{+ 3.2 }_{ -2.9 }$  &---& ${}^{+ 1.2 }_{-1.1 }$ & $\pm~ 3.0 ~$ & $\pm~ 0.7 ~$ &${}^{+ 4.4 }_{ -3.7 }$   \\
195.0 $-$ 224.0 & ${}^{+ 15.8 }_{ -14.0 }$ & ${}^{+ 3.5 }_{ -2.6 }$  &---& ${}^{+ 1.1 }_{-1.1 }$ & $\pm~ 2.8 ~$ & $\pm~ 0.7 ~$ &${}^{+ 4.9 }_{ -4.1 }$   \\
224.0 $-$ 259.0 & ${}^{+ 18.7 }_{ -16.1 }$ & ${}^{+ 4.1 }_{ -3.0 }$  &---& ${}^{+ 1.0 }_{-1.0 }$ & $\pm~ 2.6 ~$ & $\pm~ 0.6 ~$ &${}^{+ 5.8 }_{ -4.8 }$   \\
259.0 $-$ 298.0 & ${}^{+ 22.5 }_{ -18.7 }$ & ${}^{+ 4.5 }_{ -3.2 }$  &---& ${}^{+ 1.0 }_{-1.0 }$ & $\pm~ 2.4 ~$ & $\pm~ 0.7 ~$ &${}^{+ 7.0 }_{ -5.9 }$   \\
298.0 $-$ 344.0 & ${}^{+ 27.6 }_{ -22.1 }$ & ${}^{+ 5.3 }_{ -4.0 }$  &---& ${}^{+ 1.0 }_{-0.9 }$ & $\pm~ 2.3 ~$ & $\pm~ 1.2 ~$ &${}^{+ 8.7 }_{ -7.4 }$   \\
344.0 $-$ 396.0 & ${}^{+ 34.4 }_{ -26.4 }$ & ${}^{+ 5.9 }_{ -4.8 }$  &---& ${}^{+ 0.9 }_{-0.9 }$ & $\pm~ 2.2 ~$ & $\pm~ 1.7 ~$ &${}^{+ 11.2 }_{ -9.6 }$  \\
396.0 $-$ 457.0 & ${}^{+ 43.4 }_{ -32.1 }$ & ${}^{+ 7.6 }_{ -6.5 }$  &---& ${}^{+ 0.9 }_{-0.8 }$ & $\pm~ 2.1 ~$ & $\pm~ 2.5 ~$ &${}^{+ 14.6 }_{ -12.7 }$ \\
457.0 $-$ 527.0 & ${}^{+ 55.4 }_{ -39.4 }$ & ${}^{+ 10.3 }_{ -7.4 }$ &---& ${}^{+ 0.9 }_{-0.8 }$ & $\pm~ 2.3 ~$ & $\pm~ 4.6 ~$ &${}^{+ 19.5 }_{ -17.1 }$ \\
\end{tabular}
\label{tab:sys3}
\end{ruledtabular}
\end{table*}

\begin{table*}[p]
\caption{Systematic uncertainties on the measured
  inclusive jet differential cross sections
  from different sources (as discussed in Sec.~\ref{sec:Systematics})
  as a function of \pt\ for jets in the region $1.1<|y|<1.6$.}
\begin{ruledtabular}
\begin{tabular}{cccccccc}
\multicolumn{8}{c}{\normalsize{Systematic uncertainties [$\%$]
    $(1.1<|y|<1.6)$}} \\ \hline
\multicolumn{2}{c}{} &
\multicolumn{2}{c}{dijet $p_T$ balance} &
\multicolumn{4}{c}{} \\
\multicolumn{1}{c}{\pt\ (GeV/$c$)} &
\multicolumn{1}{c}{jet energy scale} &
\multicolumn{1}{c}{nominal} &
\multicolumn{1}{c}{additional} &
\multicolumn{1}{c}{pileup} &
\multicolumn{1}{c}{unfolding} &
\multicolumn{1}{c}{$p_T$-spectra} &
\multicolumn{1}{c}{resolution} \\  \hline
 62.0 $-$  72.0 & ${}^{+ 8.5 }_{ -7.8 }$   & ${}^{+ 2.4 }_{ -2.0 }$  &---                       & ${}^{+ 2.1 }_{-1.9 }$ & $\pm~ 5.5 ~$ & $\pm~ 0.2 ~$  & ${}^{+ 3.0 }_{ -2.9 }$   \\
 72.0 $-$  83.0 & ${}^{+ 9.0 }_{ -8.4 }$   & ${}^{+ 2.6 }_{ -2.1 }$  &---                       & ${}^{+ 1.9 }_{-1.8 }$ & $\pm~ 5.5 ~$ & $\pm~ 0.3 ~$  & ${}^{+ 2.9 }_{ -2.9 }$   \\
 83.0 $-$  96.0 & ${}^{+ 9.6 }_{ -9.0 }$   & ${}^{+ 2.8 }_{ -2.0 }$  &---                       & ${}^{+ 1.8 }_{-1.7 }$ & $\pm~ 5.4 ~$ & $\pm~ 0.4 ~$  & ${}^{+ 2.9 }_{ -2.8 }$   \\
 96.0 $-$ 110.0 & ${}^{+ 10.5 }_{ -9.8 }$  & ${}^{+ 2.4 }_{ -2.3 }$  &---                       & ${}^{+ 1.7 }_{-1.6 }$ & $\pm~ 5.3 ~$ & $\pm~ 0.4 ~$  & ${}^{+ 2.9 }_{ -2.9 }$   \\
110.0 $-$ 127.0 & ${}^{+ 11.6 }_{ -10.8 }$ & ${}^{+ 2.5 }_{ -2.2 }$  &---                       & ${}^{+ 1.6 }_{-1.5 }$ & $\pm~ 5.2 ~$ & $\pm~ 0.5 ~$  & ${}^{+ 3.0 }_{ -3.0 }$   \\
127.0 $-$ 146.0 & ${}^{+ 13.1 }_{ -12.0 }$ & ${}^{+ 3.4 }_{ -2.7 }$  &---                       & ${}^{+ 1.5 }_{-1.5 }$ & $\pm~ 5.1 ~$ & $\pm~ 0.6 ~$  & ${}^{+ 3.2 }_{ -3.2 }$   \\
146.0 $-$ 169.0 & ${}^{+ 15.2 }_{ -13.5 }$ & ${}^{+ 3.5 }_{ -2.6 }$  &---                       & ${}^{+ 1.5 }_{-1.4 }$ & $\pm~ 4.9 ~$ & $\pm~ 0.6 ~$  & ${}^{+ 3.7 }_{ -3.6 }$   \\
169.0 $-$ 195.0 & ${}^{+ 18.0 }_{ -15.4 }$ & ${}^{+ 4.4 }_{ -3.2 }$  &---                       & ${}^{+ 1.4 }_{-1.4 }$ & $\pm~ 4.8 ~$ & $\pm~ 0.8 ~$  & ${}^{+ 4.4 }_{ -4.3 }$   \\
195.0 $-$ 224.0 & ${}^{+ 21.7 }_{ -17.8 }$ & ${}^{+ 4.5 }_{ -3.6 }$  & ${}^{+ 2.3 }_{ -2.0 }$   & ${}^{+ 1.4 }_{-1.3 }$ & $\pm~ 4.6 ~$ & $\pm~ 0.2 ~$  & ${}^{+ 5.5 }_{ -5.2 }$   \\
224.0 $-$ 259.0 & ${}^{+ 26.6 }_{ -20.7 }$ & ${}^{+ 5.4 }_{ -4.0 }$  & ${}^{+ 5.4 }_{ -4.5 }$   & ${}^{+ 1.4 }_{-1.3 }$ & $\pm~ 4.4 ~$ & $\pm~ 1.1 ~$  & ${}^{+ 7.1 }_{ -6.6 }$   \\
259.0 $-$ 298.0 & ${}^{+ 33.1 }_{ -24.6 }$ & ${}^{+ 5.9 }_{ -5.0 }$  & ${}^{+ 9.4 }_{ -8.7 }$   & ${}^{+ 1.4 }_{-1.3 }$ & $\pm~ 4.2 ~$ & $\pm~ 0.2 ~$  & ${}^{+ 9.5 }_{ -8.6 }$   \\
298.0 $-$ 344.0 & ${}^{+ 41.8 }_{ -29.4 }$ & ${}^{+ 8.0 }_{ -5.8 }$  & ${}^{+ 16.2 }_{ -14.5 }$ & ${}^{+ 1.4 }_{-1.3 }$ & $\pm~ 4.0 ~$ & $\pm~ 3.3 ~$  & ${}^{+ 12.9 }_{ -11.4 }$ \\
344.0 $-$ 396.0 & ${}^{+ 53.4 }_{ -35.6 }$ & ${}^{+ 9.8 }_{ -6.5 }$  & ${}^{+ 29.6 }_{ -21.8 }$ & ${}^{+ 1.4 }_{-1.3 }$ & $\pm~ 3.8 ~$ & $\pm~ 10.2 ~$ & ${}^{+ 17.7 }_{ -15.3 }$ \\
396.0 $-$ 457.0 & ${}^{+ 68.8 }_{ -43.5 }$ & ${}^{+ 12.4 }_{ -7.6 }$ & ${}^{+ 52.2 }_{ -26.5 }$ & ${}^{+ 1.4 }_{-1.2 }$ & $\pm~ 3.5 ~$ & $\pm~ 21.7 ~$ & ${}^{+ 24.4 }_{ -20.7 }$ \\
\end{tabular}
\label{tab:sys4}
\end{ruledtabular}
\end{table*}

\begin{table*}[p]
\caption{Systematic uncertainties on the measured
  inclusive jet differential cross sections
  from different sources (as discussed in Sec.~\ref{sec:Systematics})
  as a function of \pt\ for jets in the region $1.6<|y|<2.1$.}
\begin{ruledtabular}
\begin{tabular}{cccccccc}
\multicolumn{8}{c}{\normalsize{Systematic uncertainties [$\%$]
    $(1.6<|y|<2.1)$}} \\ \hline
\multicolumn{2}{c}{} &
\multicolumn{2}{c}{dijet $p_T$ balance} &
\multicolumn{4}{c}{} \\
\multicolumn{1}{c}{\pt\ (GeV/$c$)} &
\multicolumn{1}{c}{jet energy scale} &
\multicolumn{1}{c}{nominal} &
\multicolumn{1}{c}{additional} &
\multicolumn{1}{c}{pileup} &
\multicolumn{1}{c}{unfolding} &
\multicolumn{1}{c}{$p_T$-spectra} &
\multicolumn{1}{c}{resolution} \\  \hline
 62.0 $-$  72.0 & ${}^{+ 9.9 }_{ -9.0 }$   & ${}^{+ 2.6 }_{ -2.3 }$   &---                       & ${}^{+ 2.5 }_{-2.4 }$ & $\pm~ 4.1 ~$  & $\pm~ 0.2 ~$ & ${}^{+ 2.5 }_{ -2.0 }$  \\
 72.0 $-$  83.0 & ${}^{+ 10.8 }_{ -9.9 }$  & ${}^{+ 2.8 }_{ -2.5 }$   &---                       & ${}^{+ 2.2 }_{-2.1 }$ & $\pm~ 3.7 ~$  & $\pm~ 0.1 ~$ & ${}^{+ 2.5 }_{ -2.0 }$  \\
 83.0 $-$  96.0 & ${}^{+ 12.0 }_{ -11.0 }$ & ${}^{+ 3.5 }_{ -2.6 }$   &---                       & ${}^{+ 2.0 }_{-1.9 }$ & $\pm~ 3.3 ~$  & $\pm~ 0.2 ~$ & ${}^{+ 2.5 }_{ -2.0 }$  \\
 96.0 $-$ 110.0 & ${}^{+ 13.8 }_{ -12.3 }$ & ${}^{+ 3.4 }_{ -2.8 }$   &---                       & ${}^{+ 1.9 }_{-1.9 }$ & $\pm~ 2.9 ~$  & $\pm~ 0.3 ~$ & ${}^{+ 2.6 }_{ -2.0 }$  \\
110.0 $-$ 127.0 & ${}^{+ 16.1 }_{ -14.1 }$ & ${}^{+ 3.6 }_{ -3.2 }$   &---                       & ${}^{+ 1.9 }_{-1.8 }$ & $\pm~ 2.7 ~$  & $\pm~ 0.3 ~$ & ${}^{+ 2.8 }_{ -2.2 }$  \\
127.0 $-$ 146.0 & ${}^{+ 19.3 }_{ -16.4 }$ & ${}^{+ 4.3 }_{ -3.8 }$   &---                       & ${}^{+ 1.9 }_{-1.9 }$ & $\pm~ 2.7 ~$  & $\pm~ 0.3 ~$ & ${}^{+ 3.2 }_{ -2.5 }$  \\
146.0 $-$ 169.0 & ${}^{+ 23.8 }_{ -19.5 }$ & ${}^{+ 4.7 }_{ -3.5 }$   &---                       & ${}^{+ 2.0 }_{-1.9 }$ & $\pm~ 2.9 ~$  & $\pm~ 0.2 ~$ & ${}^{+ 3.8 }_{ -3.0 }$  \\
169.0 $-$ 195.0 & ${}^{+ 30.0 }_{ -23.5 }$ & ${}^{+ 6.2 }_{ -6.6 }$   & ${}^{+ 4.9 }_{ -5.1 }$   & ${}^{+ 2.1 }_{-2.1 }$ & $\pm~ 3.7 ~$  & $\pm~ 0.3 ~$ & ${}^{+ 4.8 }_{ -3.8 }$  \\
195.0 $-$ 224.0 & ${}^{+ 38.1 }_{ -28.7 }$ & ${}^{+ 8.1 }_{ -5.2 }$   & ${}^{+ 10.1 }_{ -7.3 }$  & ${}^{+ 2.3 }_{-2.2 }$ & $\pm~ 5.0 ~$  & $\pm~ 0.4 ~$ & ${}^{+ 6.2 }_{ -5.0 }$  \\
224.0 $-$ 259.0 & ${}^{+ 49.2 }_{ -35.6 }$ & ${}^{+ 11.5 }_{ -7.8 }$  & ${}^{+ 20.7 }_{ -13.4 }$ & ${}^{+ 2.5 }_{-2.4 }$ & $\pm~ 7.3 ~$  & $\pm~ 1.9 ~$ & ${}^{+ 8.2 }_{ -6.7 }$  \\
259.0 $-$ 298.0 & ${}^{+ 64.1 }_{ -44.7 }$ & ${}^{+ 14.8 }_{ -11.5 }$ & ${}^{+ 27.9 }_{ -19.7 }$ & ${}^{+ 2.8 }_{-2.5 }$ & $\pm~ 11.0 ~$ & $\pm~ 7.4 ~$ & ${}^{+ 11.1 }_{ -9.2 }$ \\
\end{tabular}
\label{tab:sys5}
\end{ruledtabular}
\end{table*}

\clearpage

\begin{table*}[p]
\caption{Measured inclusive jet cross sections as a function of $p_T$
  for jets in the region $|y|<0.1$ together with the statistical
  ({\it stat.}) and systematic ({\it sys.}) uncertainties. The
  bin-by-bin parton-to-hadron-level ($C_{p\to h}$) corrections are
  also shown.}
\begin{ruledtabular}
\begin{tabular}{ccc} 
\multicolumn{1}{c}{} &
\multicolumn{1}{c}{\normalsize{$|y|<0.1     $}} &
\multicolumn{1}{c}{} \\ \hline
\multicolumn{1}{c}{$p_T$} &
\multicolumn{1}{c}{$\sigma \pm {  (stat.)} \pm {  (sys.)}$} &
\multicolumn{1}{c}{$C_{p\to h}$} \\
\multicolumn{1}{c}{(GeV/$c$)} &
\multicolumn{1}{c}{[nb/(GeV/$c$)]} &
\multicolumn{1}{c}{} \\ \hline
$  62- 72 $ & $(6.68\pm0.12^{+0.61}_{-0.58})\times 10^{0}$ & $1.072 \pm 0.107 $ \\
$  72- 83 $ & $(2.95\pm0.06^{+0.28}_{-0.27})\times 10^{0}$ & $1.054 \pm 0.086 $ \\
$  83- 96 $ & $(1.21\pm0.04^{+0.12}_{-0.11})\times 10^{0}$ & $1.040 \pm 0.069 $ \\
$  96-110 $ & $(5.44\pm0.24^{+0.57}_{-0.54})\times 10^{-1}$ & $1.028 \pm 0.055 $ \\
$ 110-127 $ & $(2.28\pm0.03^{+0.25}_{-0.23})\times 10^{-1}$ & $1.020 \pm 0.043 $ \\
$ 127-146 $ & $(9.18\pm0.14^{+1.11}_{-1.02})\times 10^{-2}$ & $1.013 \pm 0.033 $ \\
$ 146-169 $ & $(3.76\pm0.05^{+0.49}_{-0.44})\times 10^{-2}$ & $1.007 \pm 0.024 $ \\
$ 169-195 $ & $(1.38\pm0.03^{+0.20}_{-0.18})\times 10^{-2}$ & $1.003 \pm 0.017 $ \\
$ 195-224 $ & $(5.30\pm0.28^{+0.83}_{-0.73})\times 10^{-3}$ & $1.000 \pm 0.012 $ \\
$ 224-259 $ & $(1.84\pm0.04^{+0.32}_{-0.28})\times 10^{-3}$ & $0.998 \pm 0.008 $ \\
$ 259-298 $ & $(5.93\pm0.11^{+1.14}_{-0.99})\times 10^{-4}$ & $0.996 \pm 0.004 $ \\
$ 298-344 $ & $(1.75\pm0.06^{+0.38}_{-0.33})\times 10^{-4}$ & $0.996 \pm 0.002 $ \\
$ 344-396 $ & $(5.06\pm0.26^{+1.22}_{-1.04})\times 10^{-5}$ & $0.996 \pm 0.000 $ \\
$ 396-457 $ & $(1.24\pm0.11^{+0.34}_{-0.29})\times 10^{-5}$ & $0.996 \pm 0.001 $ \\
$ 457-527 $ & $(2.81\pm0.53^{+0.92}_{-0.79})\times 10^{-6}$ & $0.997 \pm 0.002 $ \\
$ 527-700 $ & $(1.81\pm0.81^{+0.78}_{-0.68})\times 10^{-7}$ & $1.000 \pm 0.001 $ \\
\end{tabular}
\end{ruledtabular}
\label{tab:sigma1}
\end{table*}

\begin{table*}[p]
\caption{Measured inclusive jet cross sections as a function of $p_T$
  for jets in the region $0.1<|y|<0.7$ together with the statistical
  ({\it stat.}) and systematic ({\it sys.}) uncertainties. The
  bin-by-bin parton-to-hadron-level ($C_{p\to h}$) corrections are
  also shown.}
\begin{ruledtabular}
\begin{tabular}{ccc} 
\multicolumn{1}{c}{} &
\multicolumn{1}{c}{\normalsize{$  0.1<|y|<0.7$}} &
\multicolumn{1}{c}{} \\ \hline
\multicolumn{1}{c}{$p_T$} &
\multicolumn{1}{c}{$\sigma \pm {  (stat.)} \pm {  (sys.)}$} &
\multicolumn{1}{c}{$C_{p\to h}$} \\
\multicolumn{1}{c}{(GeV/$c$)} &
\multicolumn{1}{c}{[nb/(GeV/$c$)]} &
\multicolumn{1}{c}{} \\ \hline
$  62- 72 $ & $(6.28\pm0.04^{+0.59}_{-0.56})\times 10^{0}$ & $1.072 \pm 0.108 $ \\
$  72- 83 $ & $(2.70\pm0.02^{+0.26}_{-0.25})\times 10^{0}$ & $1.055 \pm 0.088 $ \\
$  83- 96 $ & $(1.15\pm0.01^{+0.11}_{-0.11})\times 10^{0}$ & $1.041 \pm 0.071 $ \\
$  96-110 $ & $(4.88\pm0.03^{+0.51}_{-0.48})\times 10^{-1}$ & $1.030 \pm 0.057 $ \\
$ 110-127 $ & $(2.07\pm0.01^{+0.22}_{-0.21})\times 10^{-1}$ & $1.022 \pm 0.045 $ \\
$ 127-146 $ & $(8.50\pm0.04^{+0.98}_{-0.91})\times 10^{-2}$ & $1.015 \pm 0.035 $ \\
$ 146-169 $ & $(3.30\pm0.01^{+0.41}_{-0.38})\times 10^{-2}$ & $1.010 \pm 0.027 $ \\
$ 169-195 $ & $(1.24\pm0.01^{+0.17}_{-0.15})\times 10^{-2}$ & $1.006 \pm 0.020 $ \\
$ 195-224 $ & $(4.55\pm0.05^{+0.67}_{-0.61})\times 10^{-3}$ & $1.003 \pm 0.014 $ \\
$ 224-259 $ & $(1.56\pm0.01^{+0.25}_{-0.23})\times 10^{-3}$ & $1.002 \pm 0.010 $ \\
$ 259-298 $ & $(4.94\pm0.06^{+0.91}_{-0.80})\times 10^{-4}$ & $1.001 \pm 0.006 $ \\
$ 298-344 $ & $(1.42\pm0.02^{+0.30}_{-0.26})\times 10^{-4}$ & $1.000 \pm 0.003 $ \\
$ 344-396 $ & $(3.53\pm0.08^{+0.85}_{-0.73})\times 10^{-5}$ & $1.001 \pm 0.001 $ \\
$ 396-457 $ & $(6.87\pm0.35^{+1.93}_{-1.64})\times 10^{-6}$ & $1.001 \pm 0.000 $ \\
$ 457-527 $ & $(1.22\pm0.13^{+0.40}_{-0.34})\times 10^{-6}$ & $1.003 \pm 0.001 $ \\
$ 527-700 $ & $(7.08\pm1.97^{+3.09}_{-2.54})\times 10^{-8}$ & $1.005 \pm 0.001 $ \\
\end{tabular}
\end{ruledtabular}
\label{tab:sigma0}
\end{table*}

\begin{table*}[p]
\caption{Measured inclusive jet cross sections as a function of $p_T$
  for jets in the region $0.7<|y|<1.1$ together with the statistical
  ({\it stat.}) and systematic ({\it sys.}) uncertainties. The
  bin-by-bin parton-to-hadron-level ($C_{p\to h}$) corrections are
  also shown.}
\begin{ruledtabular}
\begin{tabular}{ccc} 
\multicolumn{1}{c}{} &
\multicolumn{1}{c}{\normalsize{$0.7<|y|<1.1$}} &
\multicolumn{1}{c}{} \\ \hline
\multicolumn{1}{c}{$p_T$} &
\multicolumn{1}{c}{$\sigma \pm {  (stat.)} \pm {  (sys.)}$} &
\multicolumn{1}{c}{$C_{p\to h}$} \\
\multicolumn{1}{c}{(GeV/$c$)} &
\multicolumn{1}{c}{[nb/(GeV/$c$)]} &
\multicolumn{1}{c}{} \\ \hline
$  62- 72 $ & $(5.32\pm0.05^{+0.55}_{-0.51})\times 10^{0}$ & $1.061 \pm 0.098 $ \\
$  72- 83 $ & $(2.32\pm0.02^{+0.24}_{-0.23})\times 10^{0}$ & $1.048 \pm 0.081 $ \\
$  83- 96 $ & $(9.83\pm0.13^{+1.06}_{-0.98})\times 10^{-1}$ & $1.038 \pm 0.067 $ \\
$  96-110 $ & $(3.95\pm0.03^{+0.43}_{-0.41})\times 10^{-1}$ & $1.031 \pm 0.055 $ \\
$ 110-127 $ & $(1.62\pm0.01^{+0.19}_{-0.17})\times 10^{-1}$ & $1.024 \pm 0.046 $ \\
$ 127-146 $ & $(6.34\pm0.04^{+0.79}_{-0.73})\times 10^{-2}$ & $1.019 \pm 0.037 $ \\
$ 146-169 $ & $(2.37\pm0.02^{+0.32}_{-0.29})\times 10^{-2}$ & $1.015 \pm 0.030 $ \\
$ 169-195 $ & $(8.41\pm0.09^{+1.27}_{-1.15})\times 10^{-3}$ & $1.012 \pm 0.024 $ \\
$ 195-224 $ & $(2.81\pm0.07^{+0.48}_{-0.43})\times 10^{-3}$ & $1.010 \pm 0.020 $ \\
$ 224-259 $ & $(8.78\pm0.10^{+1.78}_{-1.52})\times 10^{-4}$ & $1.008 \pm 0.016 $ \\
$ 259-298 $ & $(2.35\pm0.08^{+0.57}_{-0.47})\times 10^{-4}$ & $1.007 \pm 0.013 $ \\
$ 298-344 $ & $(5.37\pm0.17^{+1.59}_{-1.28})\times 10^{-5}$ & $1.007 \pm 0.011 $ \\
$ 344-396 $ & $(9.30\pm0.54^{+3.42}_{-2.67})\times 10^{-6}$ & $1.006 \pm 0.010 $ \\
$ 396-457 $ & $(1.35\pm0.18^{+0.63}_{-0.48})\times 10^{-6}$ & $1.007 \pm 0.009 $ \\
$ 457-527 $ & $(1.63\pm0.55^{+0.98}_{-0.72})\times 10^{-7}$ & $1.007 \pm 0.008 $ \\
\end{tabular}
\end{ruledtabular}
\label{tab:sigma2}
\end{table*}

\begin{table*}[p]
\caption{Measured inclusive jet cross sections as a function of $p_T$
  for jets in the region $1.1<|y|<1.6$ together with the statistical
  ({\it stat.}) and systematic ({\it sys.}) uncertainties. The
  bin-by-bin parton-to-hadron-level ($C_{p\to h}$) corrections are
  also shown.}
\begin{ruledtabular}\begin{tabular}{ccc} 
\multicolumn{1}{c}{} &
\multicolumn{1}{c}{\normalsize{$1.1<|y|<1.6$}} &
\multicolumn{1}{c}{} \\ \hline
\multicolumn{1}{c}{$p_T$} &
\multicolumn{1}{c}{$\sigma \pm {  (stat.)} \pm {  (sys.)}$} &
\multicolumn{1}{c}{$C_{p\to h}$} \\
\multicolumn{1}{c}{(GeV/$c$)} &
\multicolumn{1}{c}{[nb/(GeV/$c$)]} &
\multicolumn{1}{c}{} \\ \hline
$  62- 72 $ & $(4.57\pm0.04^{+0.50}_{-0.47})\times 10^{0}$ & $1.058 \pm 0.095 $ \\
$  72- 83 $ & $(1.81\pm0.02^{+0.21}_{-0.19})\times 10^{0}$ & $1.042 \pm 0.076 $ \\
$  83- 96 $ & $(7.39\pm0.09^{+0.88}_{-0.83})\times 10^{-1}$ & $1.029 \pm 0.059 $ \\
$  96-110 $ & $(2.97\pm0.05^{+0.37}_{-0.35})\times 10^{-1}$ & $1.020 \pm 0.046 $ \\
$ 110-127 $ & $(1.13\pm0.01^{+0.15}_{-0.14})\times 10^{-1}$ & $1.013 \pm 0.035 $ \\
$ 127-146 $ & $(4.09\pm0.03^{+0.61}_{-0.56})\times 10^{-2}$ & $1.007 \pm 0.025 $ \\
$ 146-169 $ & $(1.38\pm0.01^{+0.23}_{-0.21})\times 10^{-2}$ & $1.003 \pm 0.017 $ \\
$ 169-195 $ & $(4.13\pm0.06^{+0.82}_{-0.70})\times 10^{-3}$ & $1.000 \pm 0.011 $ \\
$ 195-224 $ & $(1.15\pm0.01^{+0.27}_{-0.22})\times 10^{-3}$ & $0.998 \pm 0.006 $ \\
$ 224-259 $ & $(2.66\pm0.07^{+0.77}_{-0.61})\times 10^{-4}$ & $0.997 \pm 0.002 $ \\
$ 259-298 $ & $(5.07\pm0.56^{+1.85}_{-1.43})\times 10^{-5}$ & $0.997 \pm 0.001 $ \\
$ 298-344 $ & $(8.19\pm0.52^{+3.90}_{-2.91})\times 10^{-6}$ & $0.997 \pm 0.003 $ \\
$ 344-396 $ & $(9.36\pm1.53^{+6.10}_{-4.32})\times 10^{-7}$ & $0.998 \pm 0.005 $ \\
$ 396-457 $ & $(6.67\pm3.40^{+6.21}_{-3.98})\times 10^{-8}$ & $1.000 \pm 0.005 $ \\
\end{tabular}
\end{ruledtabular}
\label{tab:sigma3}
\end{table*}

\begin{table*}[p]
\caption{Measured inclusive jet cross sections as a function of $p_T$
  for jets in the region $1.6<|y|<2.1$ together with the statistical
  ({\it stat.}) and systematic ({\it sys.}) uncertainties. The
  bin-by-bin parton-to-hadron-level ($C_{p\to h}$) corrections are
  also shown.}
\begin{ruledtabular}
\begin{tabular}{ccc}
\multicolumn{1}{c}{} &
\multicolumn{1}{c}{\normalsize{$1.6<|y|<2.1$}} &
\multicolumn{1}{c}{} \\ \hline
\multicolumn{1}{c}{$p_T$} &
\multicolumn{1}{c}{$\sigma \pm {  (stat.)} \pm {  (sys.)}$} &
\multicolumn{1}{c}{$C_{p\to h}$} \\
\multicolumn{1}{c}{(GeV/$c$)} &
\multicolumn{1}{c}{[nb/(GeV/$c$)]} &
\multicolumn{1}{c}{} \\ \hline
$  62- 72 $ & $(2.66\pm0.03^{+0.31}_{-0.28})\times 10^{0}$ & $1.038 \pm 0.079 $ \\
$  72- 83 $ & $(1.00\pm0.01^{+0.12}_{-0.11})\times 10^{0}$ & $1.028 \pm 0.062 $ \\
$  83- 96 $ & $(3.64\pm0.06^{+0.49}_{-0.44})\times 10^{-1}$ & $1.019 \pm 0.048 $ \\
$  96-110 $ & $(1.27\pm0.01^{+0.19}_{-0.17})\times 10^{-1}$ & $1.013 \pm 0.038 $ \\
$ 110-127 $ & $(4.12\pm0.04^{+0.70}_{-0.62})\times 10^{-2}$ & $1.008 \pm 0.030 $ \\
$ 127-146 $ & $(1.15\pm0.01^{+0.23}_{-0.20})\times 10^{-2}$ & $1.004 \pm 0.024 $ \\
$ 146-169 $ & $(2.78\pm0.07^{+0.69}_{-0.56})\times 10^{-3}$ & $1.001 \pm 0.021 $ \\
$ 169-195 $ & $(5.44\pm0.10^{+1.72}_{-1.39})\times 10^{-4}$ & $1.000 \pm 0.019 $ \\
$ 195-224 $ & $(7.89\pm0.69^{+3.24}_{-2.45})\times 10^{-5}$ & $0.999 \pm 0.019 $ \\
$ 224-259 $ & $(8.41\pm0.61^{+4.69}_{-3.38})\times 10^{-6}$ & $0.998 \pm 0.019 $ \\
$ 259-298 $ & $(5.08\pm1.39^{+3.74}_{-2.68})\times 10^{-7}$ & $0.998 \pm 0.021 $ \\
\end{tabular}
\end{ruledtabular}
\label{tab:sigma4}
\end{table*}

\end{widetext}

\end{document}